%% file: main.tex
\documentclass[pra,preprintnumbers,twocolumn,superscriptaddress,showkeys]{revtex4-2}

\input{preamble}

\begin{document}

\title{Anti-distinguishability of quantum channels}

\author{Ardra Ajitha Vijayan\,\orcidlink{0000-0003-0178-2385}}

\affiliation{Institute of Physics, Slovak Academy of Sciences, Dubravska cesta 9, Bratislava 845 11, Slovakia}
\author{Nidhin Sudarsanan Ragini\,\orcidlink{0000-0003-2205-3026}}
\email{nidhin@vt.edu}
\email{nidhin.sudarsanan@outlook.com}
\affiliation{Department of Computer Science, Virginia Tech, Blacksburg, VA 24061, USA}
\affiliation{Virginia Tech Center for Quantum Information Science and Engineering, Blacksburg, VA
24061, USA}

\begin{abstract}
We study single-query perfect anti-distinguishability/exclusion tasks of finite quantum channel ensembles, primarily using associated Choi operators and process POVMs. Weak perfect anti-distinguishability requires each measurement outcome to exclude an ensemble candidate with certainty, whereas the strong form additionally requires every candidate to have an exclusion outcome occurring with non-zero probability. For pairs of channels, both coincide with perfect distinguishability, whereas larger ensembles exhibit separations. We derive Choi-operator-support-based obstructions: weak exclusion requires at least two rank-deficient Choi operators, while strong exclusion requires the support of every candidate Choi operator to be strictly contained in the joint Choi support. The common Choi support must also lie in the kernel of the PPOVM normalisation. For commuting Choi operators, this condition is sufficient for weak exclusion when the normalisation also commutes with them. Conversely, a nonzero admixture of the same noise channel into every candidate precludes weak exclusion for any probe. We apply these results to measure-and-prepare, state-space contraction, Pauli, and replacer channels. In particular, the full permutation-channel ensemble of an unlabeled POVM is strongly anti-distinguishable exactly when some POVM effect is rank-deficient, while replacer-channel exclusion reduces to state exclusion. A Choi-commuting ensemble exhibits strong exclusion with an unassisted probe although no positive-definite PPOVM normalisation permits even weak exclusion. In contrast, commuting qubit phase gates, whose Choi operators need not commute, admit exact convex-hull criteria for weak and strong exclusion, both attainable without assistance.
\end{abstract}

\keywords{
Quantum channels;
Quantum hypothesis exclusion;
Process POVMs and quantum testers;
Quantum distinguishability;
}
	
\maketitle
A relevant quantum information processing task in the characterisation of any given quantum system is to identify or distinguish between competing hypotheses regarding the different dynamics it can undergo, albeit under specific device settings \cite{Molmer2015Hypothesis}. These hypotheses may correspond to possible dynamical models, including Hamiltonians, open system noise models etc. That said,
complete identification of any realised dynamical model does not necessarily need to be the prime operational objective. In many scenarios, it suffices to certify a target model or test a specific property of the underlying dynamics. To illustrate this with existing literature, let us restrict ourselves to Hamiltonians: instead of fully learning the Hamiltonian \cite{Granade2012Robust, Wiebe2014Hamiltonian, Yu2023Robust}, it has been established that there are regimes, and specifically chosen distance measures, where Hamiltonian testing is exponentially easier \cite{Bluhm2026Hamiltonian}.  In a similar vein, Hamiltonian certification tasks are formulated so as to enable comparisons of Hamiltonians prior to fully learning them \cite{Gao2026Certification}. 

Moreover, optimal identifiability of the hypotheses models, however, may demand resources that are unavailable to an experimenter; only a single or finite queries of the device may be permitted, auxiliary systems may not be available or of inadequate dimensionality, architecture shortcomings etc. In many such scenarios, similar to the above discussion, a weaker form of conclusion proves itself to be nevertheless useful. Rather than identifying the realised dynamics, the experimenter can design an experiment to conclusively certify that a particular dynamics was not realised. To illustrate this, consider a system whose dynamics is generated by one of the several possible Hamiltonians $H_1, \dots, H_N$. Consequently,  at any fixed interrogation time $\tau$, they respectively generate the gates $U_1^{(\tau)}, \dots, U_N^{(\tau)}$ where $U_k^{(\tau)} = \exp{(-i\tau H_k)}$. If a subsequent, well-designed, measurement experiment results in  outcome statistics, finite or full, that is inadmissible under $U_i^{(\tau)}$, then a {zero-error certificate} excluding $H_i$ is provided, without necessarily identifying the realised Hamiltonian. We ask, while  allowed experimental strategies are  restricted to those which produce a single finite outcome,  whether it is possible to design an experiment such that every observed outcome excludes at least one candidate with certainty. We distinguish this requirement from the stronger condition that every candidate can be excluded by an outcome occurring with nonzero probability for some member of the ensemble.

The above scenario is an instance of a general, single-shot, quantum hypothesis exclusion task: instead of identifying the realised member of a well-characterised collection, one looks for an outcome that conclusively rules out a specific subcollection of members. In quantum information theory, the ``static'' versions of this task, for hypothesis preparations, are referred to as quantum state anti-distinguishability (AD) or exclusion. They originated from foundational studies \cite{caves2002conditions, Pusey2012Reality}
and still continue to be explored in similar and extended spirits \cite{heinosaari2024simple,Ray2026EpistemicModels, manna2026antidistinguishabilitystatesgeneralprobabilistic}, especially through quantum contextuality perspectives \cite{leifer2020noncontextuality,srikumar2026contextuality}. In addition to their foundational implications to quantum theory, studies revealing structural constraints exist in the literature \cite{bandyopadhyay2014conclusive, heinosaari2011mathematical,russo2023inner,johnston2025tight,mcirvin2024quantum, johnston2026sharpboundsperfectquantum,roy2026quantumstateexclusioncopies} albeit most being for collections of pure states, or specifically for group-generated collections of states \cite{diebra2026quantum,yao2026conclusive,hillery2025group}. Moreover, quantum resource theoretic studies have also considered AD tasks as their test beds \cite{uola2019quantifying,ragini2024higher}.

As already indicated, the exclusion task warrants an extension from preparations to dynamical models; this is crucial since the competing hypotheses often dealt with are related to the dynamics rather than preparations. These dynamical models, when represented as quantum processes, are described by quantum channels. When they can be seen as multiple-time-step incomplete processes with memory, they are described  by quantum combs \cite{Chiribella2008_PRL,Chiribella2008_EPL_supermaps,Chiribella2009_PRA_networks}.  Recent explorations in quantum channel AD include studies on unitary channels \cite{manna2026unitaries}, and on  quantum measurement channels \cite{mannaPRAmeasurments,manna-entagldstatesForMeasurements}. Asymptotic error bounds have also been explored \cite{Ji2025Converse, Ji2026Barycentric}. But the structural aspects of AD of general channel ensembles still remain comparatively unexplored: for example, which properties of the candidate processes admit or obstruct zero-error exclusion, how these conditions are related to admissible positive-operator-valued measure (PPOVM) or testers etc.

We address these structural questions by relating the supports of the candidate Choi operators to the normalisation of the PPOVM used to test them. This approach treats the channel ensemble and the choice of probe together in determining whether a single query permits perfect exclusion. The distinction between weak and strong perfect anti-distinguishability is important here: requiring every candidate to have an exclusion outcome that can occur imposes restrictions that need not follow from requiring every observed outcome to exclude some candidate. We use this distinction to identify structural obstructions and to separate perfect exclusion from perfect identification. For ensembles with mutually commuting Choi operators, the common-support restriction becomes an exact criterion for weak perfect anti-distinguishability when the PPOVM normalisation also commutes with every candidate Choi operator. Examples within this class show, however, that commutativity does not remove the dependence on the probe. Applications to specific channel families illustrate both aspects of this analysis: they identify obstructions that no testing strategy can overcome and settings in which a suitable probe enables perfect exclusion. In particular, for the full permutation-channel ensemble of an unlabeled POVM, the existence of a rank-deficient effect exactly characterises strong perfect anti-distinguishability, giving a direct criterion for conclusively excluding possible assignments of measurement labels.

The manuscript is organised as follows. Section \ref{sec:preliminaries} introduces the Choi representation of quantum channels and the PPOVM formalism for single-query tests. Section \ref{sec:anti-distinguishability} defines weak and strong perfect anti-distinguishability, relates them to perfect distinguishability, and develops ensemble constructions and structural obstructions to perfect exclusion. Section \ref{sec:measure-and-prepare} applies these obstructions to measure-and-prepare channels and characterises strong perfect anti-distinguishability for the permutation-channel ensembles of unlabeled POVMs. Section \ref{sec:state-space-contractions} examines exclusion obstructions for state-space contractions and establishes the reduction of replacer-channel exclusion to state exclusion. Section \ref{sec:choi-commuting} derives a necessary and sufficient condition for weak perfect anti-distinguishability of Choi-commuting ensembles when the PPOVM normalisation commutes with every candidate Choi operator, and illustrates its consequences for Pauli, replacer, and measure-and-prepare channels. Section \ref{sec:phase-gates} establishes exact geometric criteria for weak and strong perfect anti-distinguishability of commuting qubit phase-gate channels. Section \ref{sec:discussions} serves as a summary of the main findings and discusses the implications of the results and directions for further work, including multiple queries and memory channels. Appendix \ref{appendix-PDchannelsPairs} examines the construction of perfectly distinguishable partner channels motivated by the ensemble-extension result in Section \ref{sec:anti-distinguishability}.

\section{Preliminaries}\label{sec:preliminaries}
\subsection{Quantum channels and Choi representation}
Within the scope of this work, all Hilbert spaces are finite-dimensional. We write $\mathcal{L}(\h_A,\h_B)$ for the linear operators from the Hilbert space $\h_A$ of a quantum system $A$ to the Hilbert space $\h_B$ of a system $B$, and $\mathcal{L}(\h_A)$ when both systems are $A$. The dimension of $A$ is the dimension of the space $\h_A$ and denoted $d_A$. Consider a general quantum evolution processing an input  system $A$ to an  output system $B$. The quantum channel $\mathcal{E}_{A \rightarrow B}$, describing this evolution, is a completely positive trace preserving map from $\mathcal{L}(\h_A )$ to $\mathcal{L}(\h_B)$.
Now, w.l.o.g, choosing the computational basis $\{\ket{x}\}_{x=0}^{d_A-1}$ for $A$ and by introducing a copy system $A^{'} $, one can define the operator $\Gamma_{AA^{'}} \in \mathcal{L}(\h_A \otimes \h_{A^{'}})$,
\begin{equation}
    \Gamma_{AA^{'}} = \dyad{\Gamma}_{AA^{'}}  = \sum_{x,y =0}^{d_A-1}  \dyad{x}{y}_A \otimes \dyad{x}{y}_{A^{'}},
\end{equation}
where $\ket{\Gamma}_{AA^{'}}  = \sum_{x=0}^{d_A-1} \ket{x}_A\otimes\ket{x}_{A^{'}}$.  Now,
through the Choi-Jamio\l{}kowski isomorphism \cite{Choi1975,Jamiolkowski1972}, $\ecal_{A \rightarrow B}$ can be represented as,
\begin{equation}
    \Gamma^{\ecal}_{AB} = (\mathcal{I}_{A}\otimes\ecal_{A^{'} \rightarrow B})[\Gamma_{AA^{'}}],
\end{equation}
where $\Gamma^{\ecal}_{AB}$ is referred to as the Choi operator of $\ecal_{A \rightarrow B}$. Complete positivity and trace-preservation of $\ecal_{A \rightarrow B}$ is respectively reflected as:
\begin{align}
    \Gamma^{\ecal}_{AB} \ge 0, \text{ and } \text{Tr}_B[\Gamma^{\ecal}_{AB}] = \id_A.
\end{align}
The inverse of the isomorphism retrieves back the action of the channel on any operator $X_A$ through,
\begin{align}
    \ecal_{A \rightarrow B}[X_A] = \text{Tr}_A[ (X^{T}_A \otimes \id_B)\Gamma^{\ecal}_{AB}].
\end{align}

Having represented quantum channels through their Choi operators, we  see how one can describe an experiment designed to test a channel with a single query.

\subsection{Testing quantum channels: Process POVMs}
 When only a single query of a quantum channel $\ecal_{A \rightarrow B}$ is available, a general ``measurement'' of this channel involves implementing it locally on a  state $\xi_{AR}$, on an extended system $AR$ assisted by an auxiliary system $R$, and then measuring the evolved state by a global POVM $M$. That is,  such a  test procedure is described by the pair $(\xi_{AR}, M)$. Such procedures are holistically described by ``process POVMs'' (PPOVM in short) \cite{ziman2008process} or single-slot PPOVMs \cite{Chiribella2009_PRA_networks}. An $n$-outcome PPOVM $T$, testing a channel $\ecal_{A \rightarrow B}$, is a collection of $n$ positive operators $\{T_x \in \mathcal{L}(\h_A \otimes \h_B)\}_{x=1}^n$ that satisfies the normalisation condition,
\begin{equation}
\label{norm}
    \sum_{x=1}^n T_x = (\text{Tr}_{R}[\xi_{AR}])^T \otimes \id_B,
\end{equation}
where the transposed input marginal state $(\text{Tr}_{R}[\xi_{AR}])^T =: \xi_{A}$ is referred to as the normalisation state of the tester. The probability of observing an outcome $x$ when $\ecal_{A \rightarrow B}$ is tested using  $T$ is given by,
\begin{equation}
    p(x; T, \ecal) = \text{Tr}[T_x\Gamma_{AB}^{\ecal}].
\end{equation}
We call a PPOVM $T$ \textit{maximally-assisted} when the
input state is the maximally entangled state. Consequently, such a PPOVM will have normalisation $\frac{\id_A}{d_A}\otimes\id_B$. In contrast,  a PPOVM admitting an implementation without any auxiliary system is \textit{unassisted}.

Equipped with the PPOVM formalism, we now formulate perfect anti-distinguishability of quantum channels and investigate the conditions under which it can be achieved.

\section{Anti-distinguishability}\label{sec:anti-distinguishability}
\begin{figure}[htbp] 
    \includegraphics[width=0.5\textwidth]{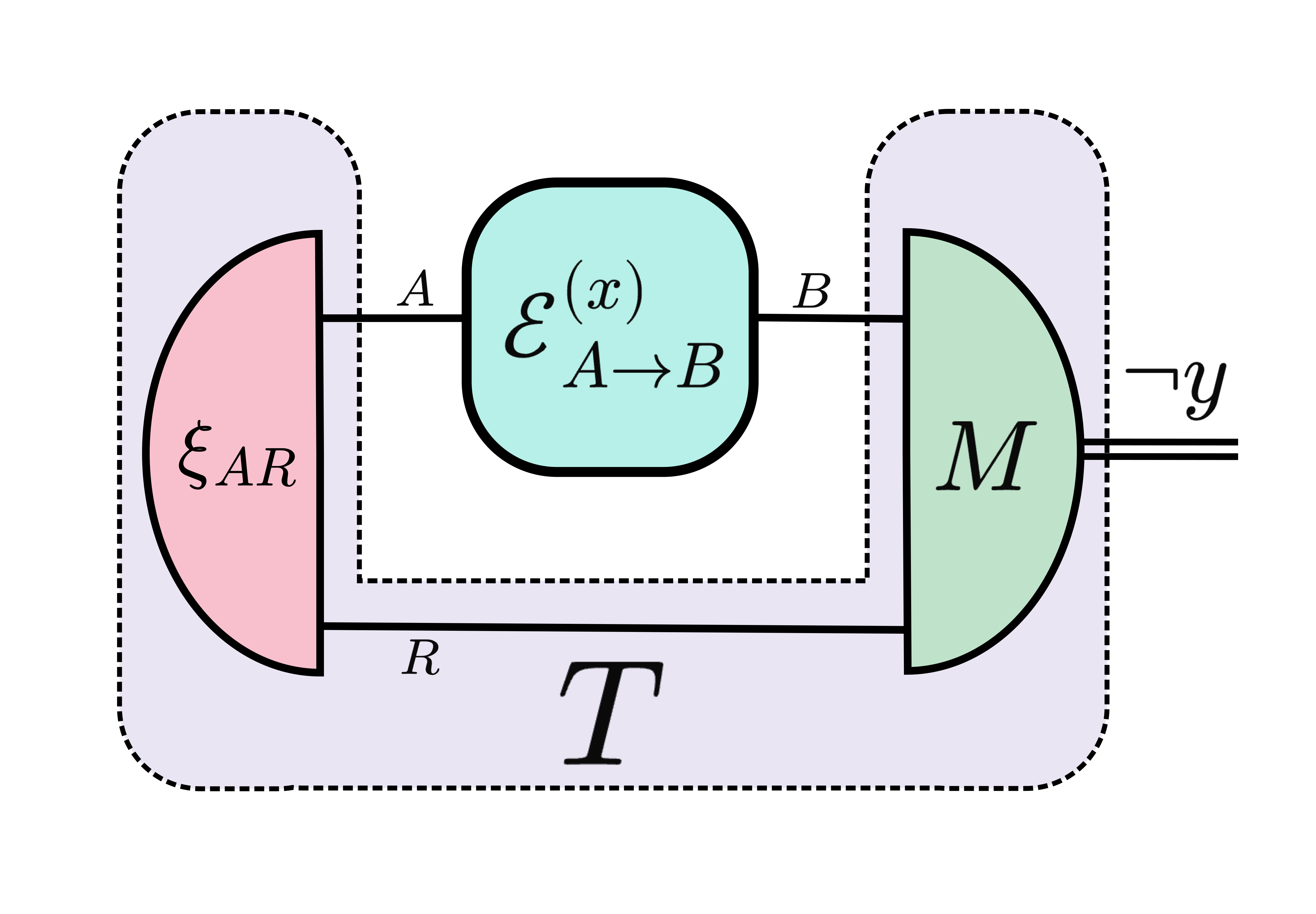} 
    \caption{Illustration of a (single-shot, single-query) quantum channel anti-distinguishability experiment; An unknown channel $\mathcal{E}_{A \rightarrow B}^{(x)}$, with label $x$, is queried from a well-characterised collection. To test this channel a probe state $\xi_{AR}$, supported on an appropriate auxiliary system $R$, can be visualised to be fed into the channel. The  output state is then measured by a POVM $M$ to register an outcome. Based on this outcome, an exclusion certificate is produced claiming that the channel is not the one labeled $y$. The pair of probe state $\xi_{AR}$ and POVM $M$ corresponds to a holistic descriptor $T$, which is known as a PPOVM.}
    \label{fig:andtidistinguishing_channels} 
\end{figure}
Let $\xcal =\{(p_i,\ecal_{A \rightarrow B}^{(i)}) \}_{i=1}^n$ be a finite ensemble of $n$ quantum channels, with  $p_i$ being the prior probability with which $\ecal_{A \rightarrow B}^{(i)}$ is chosen. Once such an ensemble is fixed for study, we suppress the channel superscript and systems subscript of involved Choi operators; the Choi operator $\Gamma_{AB}^{\ecal^{(i)}}$ of $\ecal^{(i)}_{A \rightarrow B}$ will be represented as just $\Gamma_i$. Now, in an anti-distinguishability task concerning this ensemble, an experimenter is tasked with making decisions pertaining to exclusion of each given channel. As such, the outcome set is $\{\neg x\}_{x=1}^n$ and each outcome $``\neg x"$ is associated with the PPOVM operator $T_x$. Consequently, the experiment is characterised by probabilities,
\begin{eqnarray}
    \text{Pr}[\text{guess ``not } x"|\ecal_y] &=& p(\neg x|\ecal^{(y)}_{A \rightarrow B}) \\
    &=& \text{Tr}[T_x\Gamma_y ],
\end{eqnarray}
Moreover, the probabilities with $y=x$,
\begin{equation}
    p(\neg x|\Gamma_x) = \text{Tr}[\Gamma_x T_{x}],
\end{equation}
correspond to erroneous decisions. Now, we  discuss the single-shot experimental scenario (See \autoref{fig:andtidistinguishing_channels}), and associated structures required to achieve it,  where one requires to carry out this task error-free.

\begin{definition}
    Any given ensemble $\xcal$  is said to be  \textit{weakly perfectly anti-distinguishable} (weak-PAD)  if there exists an $n$-outcome PPOVM $T$ such that,
\begin{align}
    \text{Tr}[T_x \Gamma_x] = 0 \quad \forall x. \label{eq-weak_exclusio}
\end{align} It is \textit{strongly perfectly anti-distinguishable} (strong-PAD) if, in addition to Eq.(\ref{eq-weak_exclusio}), 
\begin{align}
\sum_{k=1}^{n} \Tr[\Gamma_k T_{x}] \neq 0 \quad \forall\ x \label{eq-strong_exclusion}.
\end{align}
\end{definition}

These weak and strong versions are similar to what has been noted by Stratton et al. \cite{stratton2024operational} for state anti-distinguishability. We refer to Eq.(\ref{eq-weak_exclusio}) as ``weak-PAD" conditions; Similarly, Eq.(\ref{eq-weak_exclusio}) along with Eq.(\ref{eq-strong_exclusion}) as ``strong-PAD". Moreover, $\xcal$ is said to
admit weak(strong)-PAD if there exists a PPOVM that satisfies the weak(strong)-PAD conditions. With an abuse of usage, ``weak-PAD'' and ``strong-PAD'' are interchangeably used as a verb and as well as as adjective in this work, depending on the context.
\subsection{Structural consequences}
The weak-PAD conditions require each PPOVM effect to have vanishing overlap with its corresponding Choi operator. This imposes restrictions on the supports of the candidate Choi operators. We begin by identifying the resulting necessary conditions for weak- and strong-PAD.

\begin{proposition}[\textbf{Rank structure for AD}]\label{prop-oneChoiFullRank}
    If an ensemble is strongly-PAD, then each and every candidate Choi operator is rank-deficient. If it is weakly-PAD, then at least two of the candidates are rank-deficient.
\end{proposition}

\begin{proof}
Assume a candidate Choi operator  $\Gamma_{\alpha}$ to be  positive definite. Assume $T$ strongly anti-distinguishes the ensemble, then specifically we have $T_{\alpha} \neq 0$. But the weak PAD conditions gives
\begin{align}
    0 = \text{Tr}[T_{\alpha}\Gamma_{\alpha}] = \text{Tr}[\Gamma_{\alpha}^{1/2} T_{\alpha }\Gamma_{\alpha}^{1/2}].
\end{align}
Since $\Gamma_{\alpha}^{1/2} T_{\alpha }\Gamma_{\alpha}^{1/2} \ge 0$, vanishing trace implies $\Gamma_{\alpha}^{1/2} T_{\alpha }\Gamma_{\alpha}^{1/2} = 0$. Moreover, since $\Gamma_{\alpha} > 0$, $\Gamma_{\alpha}^{-1/2}$ exists. Sandwiching $\Gamma_{\alpha}^{1/2} T_{\alpha }\Gamma_{\alpha}^{1/2} = 0$ with $\Gamma_{\alpha}^{-1/2}$ results in $T_{\alpha} =0$, contradicting the assumption that $T$ strongly-PAD the ensemble containing $\Gamma_{\alpha}>0$. Since this holds true if any Choi operator is positive definite, for strong-PAD, each and every candidate should be rank-deficient. Now, assume $T$ only weakly anti-distinguishes the ensemble. The above argument did not use the strong-PAD condition until the final contradiction; it shows that every positive definite candidate $\Gamma_{\alpha}>0$ forces $T_{\alpha}=0$. Hence, only the process effects associated with rank-deficient candidates can be non-zero. If all candidates were positive definite, all effects would vanish, contradicting the existence of a valid PPOVM normalisation~\eqref{norm},
\begin{align}
    \sum_{x=1}^{n} T_x = \xi_A \otimes \id_B \neq 0.
\end{align}
Now, assume exactly one candidate $\Gamma_{\beta}$ to be rank-deficient. Then, $T_{x} = 0$ for all $x \neq \beta$, and the normalisation forces $T_{\beta} = \xi_A \otimes \id_B$. Using the trace-preservation condition $\text{Tr}_B[\Gamma_{\beta}] = \id_A$, together with $\text{Tr}[\xi_A] = \text{Tr}[\xi_{AR}] = 1$ (since both transposition and partial trace are trace-preserving), we obtain
\begin{align}
    \text{Tr}[T_{\beta}\Gamma_{\beta}] &= \text{Tr}[(\xi_A \otimes \id_B)\Gamma_{\beta}] \\ & = \text{Tr}\big[\xi_A \, \text{Tr}_B[\Gamma_{\beta}]\big] \\ & = \text{Tr}[\xi_A] = 1 \neq 0,
\end{align}
contradicting the weak-PAD condition $\text{Tr}[T_{\beta}\Gamma_{\beta}] = 0$. In other words, a single non-zero PPOVM effect would have to carry the whole normalisation, and it cannot then exclude its own candidate. Therefore, for weak-PAD, at least two candidates should be rank-deficient.
\end{proof}
This obstructive feature is shared by quantum channel distinguishability;  The existence of a positive definite Choi operator precludes perfect distinguishability \cite{Chiribella2008_memory_discrimination}. This warrants an exploration into how perfect distinguishability is related to perfect anti-distinguishability. For an ensemble of two channels, these two notions coincide, as shown next.

\begin{proposition}[\textbf{Equivalence for pairs}]\label{prop-dist<=>AD_for2hypo}
    For a pair of quantum channels, weak-PAD, strong-PAD and perfect distinguishability are equivalent.
\end{proposition}
\begin{proof}
    Recall that $\Gamma_1$ and $\Gamma_2$ is weakly-PAD if there exists binary PPOVM $T$, with normalisation $T_1 + T_2 = \xi_A \otimes \id_B$, such that $\text{Tr}[T_1\Gamma_1] = 0 = \text{Tr}[T_2 \Gamma_2]$. Now, using the normalisation,
    \begin{align}
        \text{Tr}[T_2 \Gamma_1] &=  \text{Tr}[(\xi_A \otimes \id_B) \Gamma_1] - \text{Tr}[T_1 \Gamma_1] \\
        &= 1-0 = 1,
    \end{align}
and similarly $\text{Tr}[T_1 \Gamma_2] =1.$ Thus, with the relabeling $D_1 = T_2, D_2 = T_1$, the PPOVM $D$ perfectly distinguishes $\Gamma_1$ and $\Gamma_2$. On the other hand, using complementary arguments, if $D$ was a PPOVM that perfectly distinguishes the pair then it weakly anti-distinguishes the pair as well. Moreover, since $\text{Tr}[T_2\Gamma_1] = \text{Tr}[T_1\Gamma_2] = 1$, every outcome occurs with nonzero probability, so that weak-PAD of a pair already implies strong-PAD. Hence, all three notions coincide.
\end{proof}
This shows that, for a pair of channels, there is no ``non-trivial'' anti-distinguishability, that it coincides with perfect distinguishability. As such, we claim that non-trivial anti-distinguishability phenomena emerge for ensembles with $N\ge 3$ channels. On the other hand, as shown next, perfect distinguishability of any number of channels implies  {strong} anti-distinguishability.
\begin{proposition}[\textbf{Consequence of perfect distinguishability}] \label{prop-PD=>PAD}
    If an ensemble $\xcal$ is perfectly distinguishable, then it is strongly-PAD. 
\end{proposition}
\begin{proof}
 Let $\xcal$ be an ensemble of $N$ channels which is  perfectly distinguished by a  PPOVM $D$. 
 One can always construct a PPOVM $T$, through a classical post-processing \cite{LeppajarviSedlak2021} of $D$, 
    \begin{equation}
        T_x = \frac{1}{(N-1)}\sum_{y\neq x} D_y.
    \end{equation}
Since $T$ is classically post-processed from $D$, both have identical normalisation. Now, note that  $\text{Tr}[D_x\Gamma_x] =1$ for all $x$ due to perfect distinguishability and consequently $\text{Tr}[D_x\Gamma_y] =\delta(x,y)$ for all $x$ and $y$. Therefore, for all $x$,
\begin{eqnarray}
    \text{Tr}[T_x \Gamma_x] &=& \frac{1}{(N-1)}\sum_{y\neq x}\text{Tr}[D_y \Gamma_x] =  0,
\end{eqnarray}
thus satisfying the weakly-PAD conditions. Now, for each $k$,
\begin{align}
    \sum_{x=1}^N \text{Tr}\left[ \Gamma_x T_k\right] &= \frac{1}{(N-1)}\sum_x \sum_{y\neq k} \text{Tr}\left[\Gamma_x D_y\right] \\
    &= \frac{1}{(N-1)}\sum_x \sum_{y \neq k} \delta(x,y) \\
    &=\frac{1}{(N-1)} \sum_{y \neq k} \sum_x\delta(x,y) \\
    &= \frac{(N-1)}{(N-1)} = 1 \neq 0,
\end{align}
thus satisfying the strongly-PAD conditions.
\end{proof}
Operationally, the above PPOVM can be implemented by first perfectly identifying the unknown channel with $D$, followed by a classical randomization that uniformly selects one of the remaining \(N-1\) candidate channels for exclusion.

Since Proposition \autoref{prop-PD=>PAD} establishes that perfect distinguishability is sufficient for strong anti-distinguishability, a natural question in the opposite direction is posed: Does strong-PAD imply perfect distinguishability? This converse, however, need not hold for ensembles with more than two channels. The following example with the Pauli gates illustrates this.

\begin{example}[\textbf{Pauli gates}]
    Let us restrict ourselves to unassisted strategies. Now, it is a well-known fact that the four Pauli gates $I,X,Y$ and $Z$ are not perfectly distinguishable without assistance. But we find that this ensemble is strongly-PAD: Use $\ket{0}$ as the probe state and then measure the  POVM $M$ with effects:
    \begin{align}
        M_I = M_Z &= \frac{1}{2} \ketbra{1}{1}, \\
        M_X = M_Y &= \frac{1}{2} \ketbra{0}{0}.
    \end{align}
        With $\beta \in \{I,X,Y,Z\}$, for each $\alpha$, we have $\text{Tr}[M_{\alpha} \beta\ketbra{0}{0}{\beta}^{\dagger}] = 0$ whenever $\alpha = \beta$. Moreover, since the outputs are $\ket{0}$ for $I, Z$ and $\ket{1}$ (up to a phase) for $X, Y$, each effect $M_{\alpha}$ has probability $\frac{1}{2}$ on the two gates of the other pair, so that every outcome occurs and the ensemble is strongly-PAD.
\end{example}

Now, we observe something peculiar from this example. If we were to partition the ensemble $\{I,X,Y,Z\}$ into two pairs $\{I,X\}$ and $\{Y,Z\}$, these pairs are individually perfectly distinguishable; both can be distinguished with the same probe $\ket{0}$ and with POVM made of $\{\ketbra{0}{0}, \ketbra{1}{1}\}$. Moreover, from Proposition \autoref{prop-dist<=>AD_for2hypo}, $\{I,X\}$ and $\{ Y,Z\}$ are individually strongly-PAD. From the above example, we found that the union ensemble $\{I,X\} \cup \{Y,Z\}$ is also strongly-PAD. Motivated by this, one can ask whether a union ensemble is perfectly anti-distinguishable if all of the sub-ensembles are individually anti-distinguishable themselves. We find that this to be in the affirmative:
\begin{proposition}[\textbf{Sub-ensembles and their unions}]\label{prop:PD_union_PAD}
    Let $\xcal_1 =\{\ecal_{A \rightarrow B}^{(i)} \}_{i=1}^n$ and $\xcal_2 =\{\ecal_{A \rightarrow B}^{(i)} \}_{i=n+1}^N$ be two ensembles such that $\xcal_1 \cap \xcal_2 = \emptyset$. 
    \begin{enumerate}
        \item  If $\mathcal{X}_1$ and $\mathcal{X}_2$ are weakly-PAD, then $\xcal_1 \cup \xcal_2$ is weakly-PAD.
        \item If $\mathcal{X}_1$ and $\mathcal{X}_2$ are strongly-PAD, then $\xcal_1 \cup \xcal_2$ is strongly-PAD.
    \end{enumerate}
\end{proposition}

\begin{proof}
Let $S$ be the PPOVM with normalisation $\omega^{(1)}_A\otimes\id_B$    that perfectly anti-distinguishes $\xcal_1$ and $R$ the one with normalisation $\omega^{(2)}_A\otimes\id_B$ that does $\xcal_2$. One can always construct an $N$-outcome PPOVM $Q$,
\begin{eqnarray}
    Q_x := \begin{cases}
        \half  S_x, &\text{ for } x= 1, \dots, n \\
        \half R_x, &\text{ for } x = n+1, \dots, N,
    \end{cases}
\end{eqnarray}

Operationally, the construction amounts to randomly selecting one of the two anti-distinguishability experiments $S$ and $R$, each with probability \(1/2\), and assigning its outcomes to the corresponding candidate labels in the union ensemble. Although the two experiments may use different probe states, their probabilistic mixture remains a valid single-query experiment, with normalisation given by the corresponding convex mixture of the original normalisation states,
\begin{eqnarray}
    \sum_{x=1}^N Q_x &=& \sum_{x=1}^n \frac{1}{2}S_x + \sum_{x=n+1}^N \frac{1}{2} R_x \\
    &=&\left(\half\omega^{(1)}_A + \half\omega^{(2)}_A\right)\otimes \id_B.
\end{eqnarray}
By virtue of $S$ and $R$, $Q_x$  weakly-PAD $\ecal_x$ for each $x = 1, \dots, n, n+1, \dots, N$:
\begin{align}
    &\text{Tr}[Q_x \Gamma_x] = \half\text{Tr}[S_x \Gamma_x] = 0 \\ &\hspace{3cm}\text{for }x = 1, \dots,n, \\
    &\text{Tr}[Q_x \Gamma_x] = \half \text{Tr}[R_x \Gamma_x] = 0 \\ & \hspace{3cm}\text{for }x = n+1, \dots,N,
\end{align}
and therefore the union $\xcal_1 \cup \xcal_2$ is weakly-PAD. Moreover, if $\xcal_1$ and $\xcal_2$ are strongly-PAD, then for $1\le k\le n$,
\begin{align}
    \sum_{x=1}^N \text{Tr}[Q_k\Gamma_x] &= \underbrace{\half \sum_{x=1}^n \text{Tr}[S_k\Gamma_x]}_{>0} + \underbrace{\half\sum_{x=n+1}^N \text{Tr}[S_k\Gamma_x]}_{\geq 0} > 0,
\end{align}
and similarly, for $n+1\le k\le N$,
\begin{align}
    \sum_{x=1}^N \text{Tr}[Q_k\Gamma_x] &= \underbrace{\half \sum_{x=1}^n \text{Tr}[R_k\Gamma_x]}_{\geq 0} + \underbrace{\half\sum_{x=n+1}^N \text{Tr}[R_k\Gamma_x]}_{>0} > 0.
\end{align}
Thus, the union is strongly-PAD as well.
\end{proof}
Equipped with this proposition, we can write down the following ``embedding" proposition as a corollary, where certain ensembles can be made PAD by combining it with appropriate ensembles if they exist.  This result is inspired from the one for quantum states, found as Proposition 2 in
\cite{heinosaari2018antidistinguishability}.

\begin{coro}[\textbf{Activation of strong-PAD}]
    For a given  ensemble $\xcal = \{ \ecal_{A \rightarrow B}^{(i)}  \}_{i=1}^n$, if there exists an ensemble $\{\fcal_{A \rightarrow B}^{(i)}\}_{i=1}^n$ such that each pair $\xcal_i = \{\ecal_{A \rightarrow B}^{(i)},\fcal_{A \rightarrow B}^{(i)}\}$ is perfectly distinguishable, then $\xcal$ is a sub-ensemble of a strong-PAD admitting ensemble ${\xcal}^{'}$ with at most $2n$ channels.
\end{coro}  

\begin{proof}
    Assume that for $\xcal = \{ \ecal_{A \rightarrow B}^{(i)}  \}_{i=1}^n $, there exists $n$ channels $\{ \fcal_{A \rightarrow B}^{(i)}\}$ such that $\{ \ecal_{A \rightarrow B}^{(i)}, \fcal_{A \rightarrow B}^{(i)}\}$ is perfectly distinguishable for each $i$. Consequently, each of these pairs are also perfectly anti-distinguishable due to Proposition \ref{prop-PD=>PAD}. Now, it can be seen that the union ensemble,
    \begin{equation}
        \xcal^{'} = \bigcup_{i}^n \xcal_i,
    \end{equation}
    {is strong-PAD} where $|{\xcal^{'}}| \le 2n$.
\end{proof}
Moreover, the operational significance  of this corollary lies in the following observation: Suppose we are given the ensemble $\xcal$ for which the desired distinguishability conditions can be satisfied. Then, we know that by appropriately ``extending'' the ensemble, strong-PAD  is activated at this cost. One can also find a special instance, for triples of qubit unitary channels, of this corollary as Theorem 4 from \cite{manna2026unitaries}.

The above corollary requires a perfectly distinguishable partner channel for each candidate of the original ensemble. This motivates us to examine when such partners can be identified, beginning with the restricted scenario where only unassisted PPOVMs are available.

\begin{example}
This example pertains to a restricted scenario where we want to identify a ``perfect distinguishability partner'' channel $\fcal_{A\rightarrow B}$ to a given channel $\ecal_{A\rightarrow B}$, while having access only to unassisted PPOVMs, such that the pair is perfectly distinguishable.

Suppose there exists a quantum state $\varrho_A$ such that the output state $\ecal_{A\rightarrow B}[\varrho_A]$ is rank-deficient. Then its kernel is nontrivial, and for any
\[
    \ket{\varphi}_B\in
    \ker\!\left(\ecal_{A\rightarrow B}[\varrho_A]\right),
\]
we have
\begin{equation}
    \ecal_{A\rightarrow B}[\varrho_A]\ket{\varphi}_B=0,
\end{equation}
or, equivalently,
\begin{equation}
    \bra{\varphi}_B
    \ecal_{A\rightarrow B}[\varrho_A]
    \ket{\varphi}_B=0.
\end{equation}

Now consider the complete state-space contraction channel
$\fcal^{(\ket{\varphi})}_{A\rightarrow B}$ defined by
\begin{align}
    \fcal^{(\ket{\varphi})}_{A\rightarrow B}[\sigma_A]
    = \dyad{\varphi}_B,
\end{align}
for every input state $\sigma_A$. In particular, when the same state $\varrho_A$ is used as an unassisted probe for the two channels, their respective output states are
\begin{equation}
    \ecal_{A\rightarrow B}[\varrho_A]
    \qquad\text{and}\qquad
    \fcal^{(\ket{\varphi})}_{A\rightarrow B}[\varrho_A]
    =\dyad{\varphi}_B.
\end{equation}
By construction,
\begin{equation}
    \ecal_{A\rightarrow B}[\varrho_A]\,
    \fcal^{(\ket{\varphi})}_{A\rightarrow B}[\varrho_A]
    =
    \ecal_{A\rightarrow B}[\varrho_A]\dyad{\varphi}_B
    =0.
\end{equation}
Thus, the two output states have orthogonal supports and can therefore be perfectly distinguished by a measurement performed solely on the output system $B$. Consequently, $\ecal_{A\rightarrow B}$ and
$\fcal^{(\ket{\varphi})}_{A\rightarrow B}$ are perfectly distinguishable using the single unassisted probe state $\varrho_A$.
\end{example}
Therefore, if we have an ensemble of channels for each of which we can identify an unassisted probe state yielding a rank-deficient output, then a corresponding complete state-space contraction channel can be constructed by choosing its output state within the kernel of that rank-deficient output. Hence, even under the restriction to unassisted strategies, one can engineer a scenario that admits PAD.

As illustrated next, more generally, we can always find a perfect distinguishability partner channel, and appropriate PPOVM normalisation, for a given channel if its Choi operator is rank-deficient.

\begin{example}[\textbf{Existence of a perfectly distinguishable partner}]
Let $\mathcal{E}_{A\to B}$ be a quantum channel whose Choi
operator $\Gamma^{\mathcal E}_{AB}$ is rank-deficient, i.e.,
$\ker(\Gamma^{\mathcal E}_{AB})\neq\{0\}$.
Define
\begin{equation}
    \Pi_{AB}:=\Pi_{\ker(\Gamma^{\mathcal E}_{AB})},
    \qquad
    \Xi_A:=\operatorname{Tr}_B[\Pi_{AB}],
\end{equation}
where $\Pi_{AB}$ is the orthogonal projector onto the
kernel of $\Gamma^{\mathcal E}_{AB}$. A perfectly distinguishable partner channel
$\mathcal{F}_{A\to B}$ can then be constructed with
Choi operator
\begin{equation}
\begin{aligned}
    \Gamma^{\mathcal F}_{AB}
    :={}&(\Xi_A^{-1/2}\otimes\mathbbm{1}_B)
    \Pi_{AB}
    (\Xi_A^{-1/2}\otimes\mathbbm{1}_B)\\
    &\hspace{2.3cm}+\frac{1}{d_B}
    P_{\ker(\Xi_A)}\otimes\mathbbm{1}_B,
\end{aligned}
\end{equation}
where $\Xi_A^{-1/2}$ denotes the Moore-Penrose
pseudoinverse square root of $\Xi_A$, and $d_B$
is the dimension of the output system $B$. The pair $\{\mathcal E_{A\to B},\mathcal F_{A\to B}\}$
is perfectly distinguishable in the single-shot
assisted scenario with the PPOVM normalisation state
\begin{equation}
    \xi_A:=
    \frac{\Xi_A^{1/2}}
    {\operatorname{Tr}[\Xi_A^{1/2}]}.
\end{equation}
In particular, the constructed operators satisfy
\begin{equation}
    \Gamma^{\mathcal E}_{AB}
    (\xi_A\otimes\mathbbm{1}_B)
    \Gamma^{\mathcal F}_{AB}=0,
\end{equation}
which is the condition for perfect distinguishability of the said candidate channels \cite{Chiribella2008_memory_discrimination}.

Appendix \ref{appendix-PDchannelsPairs} establishes the validity of the constructed
Choi operator, verifies the perfect distinguishability
condition, and proves the converse: the existence of
such a partner requires $\Gamma^{\mathcal E}_{AB}$
to be rank-deficient.
\end{example}

The preceding results describe constructions of perfectly anti-distinguishable ensembles, particularly through perfectly distinguishable channel pairs. We now identify conditions that obstruct perfect anti-distinguishability of a given ensemble. For this, we examine the common support of the candidate Choi operators together with the PPOVM normalisation.

\subsubsection{{Necessary condition for weak-PAD} }For a given ensemble $\xcal$, define the common Choi support,
\begin{equation}
    S_{\xcal} := \bigcap_{i=1}^n \text{supp}(\Gamma_i),
\end{equation}
where $\text{supp}(X)$ is the support of the operator $X$. We call $S$ the ``common Choi support'' associated to $\xcal$. A necessary condition for weak-PAD associated with $S$ is characterized as the following.
\begin{theorem}[\textbf{Choi support inclusion as a necessity}]\label{theorem-PAD_necessary} 
    If an ensemble $\xcal$  admits weak-PAD with a PPOVM whose normalisation is $\xi_A \otimes\id_B$, its common Choi support must lie entirely within the kernel of the normalisation:
    \begin{align}
        S_{\xcal} \subseteq \text{ker}(\xi_A) \otimes \h_B.
    \end{align}
\end{theorem}
\begin{proof}
    From weak-PAD conditions, we have $\text{Tr}(\Gamma_x T_x) = 0$ for all $x$. Since all $\Gamma_x$ and $T_x$ are positive operators, this essentially translates to $\Gamma_x T_x = O = T_x \Gamma_x$ for all $x$. That is, every vector in $\text{supp}(\Gamma_x)$ must lie in $\text{ker}(T_x)$, that is $\text{supp}(\Gamma_x) \subseteq \text{ker}(T_x)$. Taking intersections over $x$:
    \begin{align}
        S_{\xcal} \subseteq \bigcap_{x=1}^n \text{ker}(T_x) &= \text{ker}\left( \sum_{x=1}^n T_x\right) \\
        &= \text{ker}(\xi_A \otimes \id_B) \\
        &= \text{ker}(\xi_A) \otimes \h_B.
    \end{align}
Here, we have used the following identities. First, for two positive operators $P_1$ and $P_2$, we have $\text{ker}(P_1 + P_2)= \text{ker}(P_1) \cap \text{ker}(P_2)$. Second, for two operators $X_A,Y_B$, $\text{ker}(X_A \otimes Y_B) = \ker(X_A) \otimes \h_B + \h_A \otimes \text{ker}(Y_B) $.
\end{proof}

Theorem \ref{theorem-PAD_necessary} imposes a necessary condition that relates the common Choi support of an ensemble to the normalisation of the PPOVM used to test it. We now examine some immediate consequences of this condition, beginning with ensembles whose candidate Choi operators are all positive definite.

\begin{remark}[\textbf{Obstruction from positive definiteness}]
    Recall that an ensemble of only positive definite Choi operator channels can never be perfectly anti-distinguished. This can also be observed separately from Theorem \ref{theorem-PAD_necessary} as follows:  $S$ coincides with the full vector space $\h_A \otimes \h_B$ when all the Choi operators are positive definite. Then, the requirement from the theorem $S \subseteq \text{ker}(\xi_A) \otimes \h_B$ translates to
    \begin{align}
        \h_A \otimes \h_B \subseteq \text{ker}(\xi_A) \otimes \h_B.
    \end{align}
    This holds true only when $\text{ker}(\xi_A) $ coincides with $\h_A$, which is not admitted for any bonafide quantum state $\xi_A$. Hence, such an ensemble cannot be perfectly anti-distinguished.
\end{remark}

A further consequence of Theorem \ref{theorem-PAD_necessary} can be obtained when the PPOVM normalisation state itself is positive definite. In this case, the necessary condition reduces to a restriction involving only the common Choi support.

\begin{coro}[\textbf{Trivial Choi support for positive definite normalisation}]
    If an ensemble $\xcal$ is weakly-PAD by a PPOVM with a positive definite normalisation $\xi_A >0$, then the associated common Choi support is necessarily trivial:
    \begin{align}
        S_{\xcal} = \{ 0\}.
    \end{align}
\end{coro}
\begin{proof}
    For a positive definite $\xi_A >0$, $\text{ker}(\xi_A) = \{0\}$. Hence, from Theorem \ref{theorem-PAD_necessary},
    \begin{align}
        S_{\xcal} &\subseteq \text{ker}(\xi_A) \otimes \h_B \\
        &= \{0\} \otimes \h_B \\
        &= \{0\}.
    \end{align}
\end{proof}
\begin{example}[\textbf{Full Schmidt-rank probes}]
    Every pure probe state with full Schmidt rank will correspond to a PPOVM with positive normalisation state. An important instance of this case is when the probe states are maximally entangled, where the normalisations are always positive $\frac{\id_A}{d_A} >0$; As such, to weakly-PAD an ensemble with maximally assisted PPOVMs requires the associated common Choi support to be trivial.
\end{example}

The above condition does not rule out weak-PAD for an ensemble with nontrivial common Choi support when the PPOVM normalisation state is rank-deficient. Such a normalisation may still satisfy Theorem \ref{theorem-PAD_necessary}, as illustrated by the following example.

\begin{example}[\textbf{Maximal assistance does not help when product probes do}]
With $\omega = \exp(2\pi i /3)$, define the three qubit trine states $\ket{\psi^+_k}$ and their respective orthogonal states $\ket{\psi^-_k}$:
\begin{align}
    \ket{\psi_k^{\pm}} = \frac{1}{\sqrt{2}}(\ket{0} \pm \omega^k \ket{1}) \text{ for } k=0,1,2,
\end{align}
  and the three qubit measure-and-prepare channels with action,
  \begin{align}
      \mathcal{E}_k(\varrho) = \braket{0}{\varrho|0}\ketbra{\psi_k^+}{\psi_k^+} + \braket{1}{\varrho|1}\ketbra{0}{0}
  \end{align}
  with respective Choi operators
  \begin{align}
      \Gamma_k = \ketbra{0}{0} \otimes \ketbra{\psi_k^+}{\psi_k^+} + \ketbra{1}{1} \otimes \ketbra{0}{0}.
  \end{align}
  We have $S = \text{span}(\{ \ket{1}_A\ket{0}_B\}) \neq \{ 0\}$.
  Now, consider any PPOVM with positive normalisation state $\xi_A >0$. Then, for weakly-PAD we require
  \begin{align}
      \text{span}(\{ \ket{1}_A\ket{0}_B\}) \subseteq \{0\},
  \end{align}
  which is never true and hence such a PPOVM can never weakly-PAD the ensemble. Moreover, we can read off from this that no maximally assisted PPOVM can realise weakly-PAD for this ensemble as well. But, crucially, use a product probe state with the channel input state vector being $\ket{0}$, each of the channels output the respective trine state. Measure the POVM with effects $M_k= \frac{2}{3} \ketbra{\psi_k^{-}}{\psi_k^{-}}$. Consequently, $\text{Tr}[\ecal_k(\varrho) M_k] =0$ for all $k$, thus weakly-PAD the three channels. In fact, since each $M_k$ has probability $\frac{1}{2}$ on the other two trine states, every outcome occurs, and this product probe even achieves strong-PAD.
\end{example}
The above example also exemplifies the fact that entangled states are not always optimal for perfect anti-distinguishability, a product probe achieves strong-PAD, whereas no maximally-assisted PPOVM achieves even weak-PAD; This is similar to the exemplification by Manna et al. \cite{manna2026unitaries} in their Theorem 3, where their specific ensemble of three unitary channels admits weak-PAD without maximally entangled probes, at the same time failing to admit weak-PAD with maximally entangled probes.
The  example shows that the admissibility of weak-PAD can depend on the PPOVM normalisation. However, some ensembles do not admit weak-PAD for any normalisation state. One such instance arises when every candidate contains a nonzero contribution from a common quantum channel, as established next.

\begin{proposition}[\textbf{Classical mixing-induced inadmissibility}]\label{prop:mix-a-fixed-Choi}
    Let  $\xcal = \{ \Gamma_x\}_{x=1}^N$ be an arbitrary given ensemble and let $\Gamma^{\mathcal{N}}$ be the Choi operator of a fixed channel $\mathcal{N}_{A \rightarrow B}$. Then, any ensemble composed of physical mixtures of each and every candidate of $\xcal$ with $\Gamma^{\mathcal{N}}$,
    \begin{align}
        \tilde{\Gamma}_{x} := (1-p_x) \Gamma_x+ p_x \Gamma^{\mathcal{N}} \text{ with }p_x>0 ~\forall x,
    \end{align}
    does not admit weak-PAD.
\end{proposition} 
\begin{proof}
    Suppose there exists a PPOVM $T$ with normalisation $\xi_A \otimes\id_B$ weakly-PAD $\{ \tilde{\Gamma}_x\}_{x=1}^N$. Then, the weak-PAD conditions for all $x$ translate to
    \begin{align}
        0&= \text{Tr}[\tilde{\Gamma}_x T_x] \\
        &= (1-p_x)\text{Tr}[\Gamma_x T_x] + p_x \text{Tr}[\Gamma^{\mathcal{N}}T_x],
    \end{align}
    from which the following separately holds,
    \begin{align}
        \text{Tr}[\Gamma^{\mathcal{N}}T_x] = 0, ~~\forall x.
    \end{align}
    Summing over $x$, 
    \begin{align}
        0 &=\sum_x\text{Tr}[\Gamma^{\mathcal{N}}T_x] \\
        &= \text{Tr}\left[ \Gamma^{\mathcal{N}} \left( \sum_x T_x\right)\right] \\
        &= \text{Tr}[\Gamma^{\mathcal{N}} (\xi_A \otimes \id_B)]\\
        &= 1,
    \end{align}
    which is a contradiction and consequently we conclude that there does not exist a PPOVM normalised by $\xi_A \otimes \id_B$ that weak-PAD the said ensemble.
\end{proof}

\begin{example}[\textbf{On incompatibility-broken ensembles}]\label{example:incompatibility-broken}
    An ensemble of channels is said to be incompatible if there does not exist a multiple-output joint channel from which action of each of the candidate channel can be retrieved by selecting a specific output and discarding the rest \cite{Heinosaari2017Incompatibility}. Given any such ensemble $\{ \Gamma_x\}_{x=1}^N$, an operational way to break the incompatibility, or make the ensemble compatible, is to physically mix some appropriately chosen noise channel $\mathcal{N}_x$ (with Choi operator $\Gamma^{\mathcal{N}_x}$) to each  candidate $\Gamma_x$ such that the ensemble 
    \begin{align}
        \{ (1-\mu_x) \Gamma_x + \mu_x \Gamma^{\mathcal{N}_x}\}_{x=1}^N~~;~~ \mu_x \in (0,1] ~\forall x,
    \end{align}
    is compatible for some mixing values $\mu_x$. For instances where the chosen noise channels are identical $\Gamma^{\mathcal{N}_x} = \Gamma^{\mathcal{N}_y}$, for all $x,y$, the ensemble resembles one discussed in Proposition \ref{prop:mix-a-fixed-Choi}. Thus, we can assert that such incompatibility-broken or ``compatibilised'' ensembles do not admit weak-PAD.
\end{example}

\subsubsection{Necessary condition for strong-PAD}
Theorem \ref{theorem-PAD_necessary} imposes a necessary condition for weak-PAD. For strong-PAD, Proposition \ref{prop-oneChoiFullRank} additionally requires each and every candidate Choi operator to be rank-deficient. This requirement can be sharpened by comparing the support of each candidate Choi operator with the subspace jointly spanned by the supports of all the candidates, rather than with the full space $\h_A \otimes \h_B$. For a given ensemble $\xcal$, define the joint Choi support,
\begin{equation}
    J_{\xcal} := \sum_{i=1}^n \text{supp}(\Gamma_i) = \text{supp}(\Gamma_{\xcal}), \quad \text{with } \Gamma_{\xcal} := \sum_{i=1}^n \Gamma_i,
    \label{eq:joint-choi-support}
\end{equation}
where the sum of subspaces denotes their span, and the second equality holds since all $\Gamma_i \ge 0$. While the common Choi support $S_{\xcal}$ collects the vectors shared by each and every candidate, the joint Choi support $J_{\xcal}$ is the smallest subspace containing the supports of all the candidates. A necessary condition for strong-PAD associated with $J_{\xcal}$ is characterised as the following.

\begin{proposition}[\textbf{Strict joint Choi support inclusion as a necessity}]\label{prop:strong-pad-joint-support}
    If an ensemble $\xcal$ admits strong-PAD, then the support of each and every candidate Choi operator is strictly contained in the joint Choi support:
    \begin{align}
        \text{supp}(\Gamma_x) \subsetneq J_{\xcal} \quad \forall x.
        \label{eq:joint-support-strict}
    \end{align}
    Equivalently, $\text{rank}(\Gamma_x) < \dim(J_{\xcal})$ for all $x$, or $Q_x \Gamma_{\xcal} Q_x \neq 0$ for all $x$, where $Q_x$ is the orthogonal projector onto $\text{ker}(\Gamma_x)$.
\end{proposition}
\begin{proof}
    By definition, $\text{supp}(\Gamma_x) \subseteq J_{\xcal}$, or equivalently $\text{ker}(\Gamma_{\xcal}) \subseteq \text{ker}(\Gamma_x)$, for every $x$. Therefore, it suffices to rule out equality.

    Now, assume $T$, with normalisation $\xi_A \otimes \id_B$, strongly anti-distinguishes $\xcal$. As in the proof of Theorem \ref{theorem-PAD_necessary}, since $T_x$ and $\Gamma_x$ are positive operators, the weak-PAD conditions Eq.(\ref{eq-weak_exclusio}) translate to $\Gamma_x T_x = 0$ for all $x$. That is,
    \begin{align}
        \text{supp}(T_x) \subseteq \text{ker}(\Gamma_x) \quad \forall x.
        \label{eq:Tx-in-kernel}
    \end{align}
    Moreover, the strong-PAD condition Eq.(\ref{eq-strong_exclusion}) reads
    \begin{align}
        \text{Tr}[\Gamma_{\xcal} T_x] = \sum_{k=1}^n \text{Tr}[\Gamma_k T_x] > 0 \quad \forall x.
        \label{eq:sum-nonzero}
    \end{align}
    Assume $\text{supp}(\Gamma_{\alpha}) = J_{\xcal}$ for some candidate label $\alpha$. Then, $\text{ker}(\Gamma_{\alpha}) = J_{\xcal}^{\perp} = \text{ker}(\Gamma_{\xcal})$, and Eq.(\ref{eq:Tx-in-kernel}) gives $\text{supp}(T_{\alpha}) \subseteq \text{ker}(\Gamma_{\xcal})$. This results in $\Gamma_{\xcal} T_{\alpha} = 0$ and hence $\text{Tr}[\Gamma_{\xcal} T_{\alpha}] = 0$, contradicting Eq.(\ref{eq:sum-nonzero}). Therefore, the inclusion must be strict for each and every candidate.

For the equivalent forms, recall that $\text{supp}(\Gamma_x) \subseteq J_{\xcal}$. For two subspaces, one contained in the other, the inclusion is strict if and only if the smaller one has strictly lower dimension. Since $\text{rank}(\Gamma_x) = \dim \text{supp}(\Gamma_x)$, this gives the rank form. 

Moreover, $Q_x \Gamma_{\xcal} Q_x = A^{\dagger} A$ with $A := \Gamma_{\xcal}^{1/2} Q_x$, which vanishes if and only if $A = 0$. Since $Q_x$ projects onto $\text{ker}(\Gamma_x)$ and $\text{ker}(\Gamma_{\xcal}^{1/2}) = \text{ker}(\Gamma_{\xcal})$, the condition $A = 0$ is equivalent to $\text{ker}(\Gamma_x) \subseteq \text{ker}(\Gamma_{\xcal})$. Combined with the reverse inclusion $\text{ker}(\Gamma_{\xcal}) \subseteq \text{ker}(\Gamma_x)$ noted above, this is equivalent to $\text{ker}(\Gamma_x) = \text{ker}(\Gamma_{\xcal})$, that is, $\text{supp}(\Gamma_x) = J_{\xcal}$. Hence $Q_x \Gamma_{\xcal} Q_x \neq 0$ if and only if $\text{supp}(\Gamma_x) \subsetneq J_{\xcal}$. 
\end{proof}
Operationally, if the support of a candidate $\Gamma_{\alpha}$ coincides with the joint Choi support, then any PPOVM effect that conclusively excludes $\ecal^{(\alpha)}_{A \rightarrow B}$ is supported where no candidate responds. Such an effect has vanishing probability for each and every member of the ensemble, and the outcome ``$\neg \alpha$'' never occurs. Since $J_{\xcal} \subseteq \h_A \otimes \h_B$, Proposition \ref{prop:strong-pad-joint-support} implies the strong-PAD part of Proposition \ref{prop-oneChoiFullRank}, and the two conditions are equivalent when $J_{\xcal} = \h_A \otimes \h_B$. In general, however, $J_{\xcal}$ need not be the full space, and as illustrated next, Proposition \ref{prop:strong-pad-joint-support} is then strictly stronger: rank-deficiency of each and every candidate Choi operator does not guarantee strong-PAD.

\begin{example}[\textbf{Rank-deficient candidates without strong-PAD}]\label{ex:weak-not-strong-noncommuting}
    Consider qubit channels, that is, $d_A = d_B = 2$. With $\sigma_0 = \id$, $\sigma_X = X$, $\sigma_Y = Y$ and $\sigma_Z = Z$, define the normalised Bell vectors
    \begin{align}
        \ket{r_{\alpha}} := \frac{1}{\sqrt{2}} (\id_A \otimes \sigma_{\alpha}) \ket{\Gamma}_{AB}, \quad \alpha \in \{0, X, Y, Z\},
        \label{eq:bell-vectors}
    \end{align}
    which form an orthonormal basis of $\h_A \otimes \h_B$. For a qubit unitary $U$, the unitary channel $\mathcal{U}^{U}_{A \rightarrow B}[\varrho] := U \varrho U^{\dagger}$ has the Choi operator
    \begin{align}
        \Gamma^{\mathcal{U}^{U}}_{AB} = (\id_A \otimes U)\dyad{\Gamma}_{AB}(\id_A \otimes U^{\dagger}) = 2\dyad{r_U},
    \end{align}
    where $\ket{r_U} := \frac{1}{\sqrt{2}}(\id_A \otimes U)\ket{\Gamma}_{AB}$. Now, let $H = (X+Z)/\sqrt{2}$ be the Hadamard gate, so that $\ket{r_H} = (\ket{r_X} + \ket{r_Z})/\sqrt{2}$, and consider the ensemble $\xcal = \{\ecal^{(i)}_{A \rightarrow B}\}_{i=1}^3$ with
    \begin{align}
        \ecal^{(1)}_{A \rightarrow B} &= \half \mathcal{U}^{H}_{A \rightarrow B} + \half \mathcal{U}^{Z}_{A \rightarrow B}, \\
        \ecal^{(2)}_{A \rightarrow B} &= \mathcal{U}^{X}_{A \rightarrow B}, \\
        \ecal^{(3)}_{A \rightarrow B} &= \mathcal{U}^{Z}_{A \rightarrow B}.
        \label{eq:hadamard-ensemble}
    \end{align}
    The respective Choi operators are
    \begin{align}
        \Gamma_1 &= \dyad{r_H} + \dyad{r_Z}, \\
        \Gamma_2 &= 2\dyad{r_X}, \\
        \Gamma_3 &= 2\dyad{r_Z}.
    \end{align}
    Since $\ket{r_H}$ is a coherent superposition of $\ket{r_X}$ and $\ket{r_Z}$, we have $\text{supp}(\Gamma_1) = \text{span}\{\ket{r_X}, \ket{r_Z}\}$. Hence $\text{rank}(\Gamma_1) = 2$ and $\text{rank}(\Gamma_2) = \text{rank}(\Gamma_3) = 1$; that is, each and every candidate is rank-deficient, and Proposition \ref{prop-oneChoiFullRank} does not obstruct strong-PAD. However, the joint Choi support is $J_{\xcal} = \text{span}\{\ket{r_X}, \ket{r_Z}\} = \text{supp}(\Gamma_1)$. From Proposition \ref{prop:strong-pad-joint-support}, the ensemble does not admit strong-PAD for any PPOVM normalisation, assisted or unassisted.

    Nevertheless, the ensemble admits weak-PAD. Consider the maximally-assisted PPOVM with effects
    \begin{align}
        T_1 &= \half \left( \dyad{r_0} + \dyad{r_Y} \right), \\
        T_2 &= \half \dyad{r_Z}, \\
        T_3 &= \half \dyad{r_X}.
    \end{align}
    These effects have the normalisation $\sum_x T_x = \frac{\id_A}{2} \otimes \id_B$ and satisfy $\text{Tr}[\Gamma_x T_x] = 0$ for $x = 1,2,3$. Operationally, one feeds half of a maximally entangled state into the unknown channel and then measures the output in the Bell basis. In accordance with the failure of strong-PAD, $\sum_{k} \text{Tr}[\Gamma_k T_1] = 0$ even though $T_1 \neq 0$: every observed outcome excludes either $\ecal^{(2)}_{A \rightarrow B}$ or $\ecal^{(3)}_{A \rightarrow B}$, while $\ecal^{(1)}_{A \rightarrow B}$ is never excluded. Thus, rank-deficiency of each and every candidate is not sufficient for strong-PAD.
\end{example}
The preceding results and discussions provide necessary conditions for perfect anti-distinguishability of general  ensembles, together with constructions of ensembles admitting perfect exclusion. We now apply these structural results to study specific families of non-unitary channels, beginning with measure-and-prepare channels.
\section{Measure-and-prepare channels}\label{sec:measure-and-prepare}
An $n$-outcome quantum measurement $M$,  described as a POVM, can also be modeled as a quantum channel. Referred to as  the quantum-to-classical channel of $M$, its action is given by
\begin{align}
    \mathcal{M}^M(\varrho) = \sum_{k=1}^n \text{Tr}[\varrho M_k] \dyad{k}.
\end{align}
For a given POVM $M$, a whole family of such channels can be constructed, so as to describe more general scenarios of measurement-conditioned state preparations. Befittingly, they are referred to as measure-and-prepare channels \cite{HorodeckiShorRuskai2003EntanglementBreaking}, whose formal definition is the following.
Given  $M$ and a ``preparation set'' of $n$ states $\mathcal{S} =\{ \varsigma_k\}_{k=1}^n$, associated measure-and-prepare channel $\mcal^{M,\mathcal{S}}$ is defined through its action as,
\begin{equation}
    \mcal^{M,\mathcal{S}}(\varrho) = \sum_k\text{Tr}[M_k \varrho]\varsigma_k.
\end{equation}
Its Choi operator is given by,
\begin{equation}
    \Gamma^{M,\mathcal{S}}_{AB} = \sum_k M_k^{T} \otimes \varsigma_k.
\end{equation}

We now examine when a measure-and-prepare channel can obstruct strong-PAD. From Proposition \ref{prop-oneChoiFullRank}, a positive definite Choi operator is sufficient to preclude strong-PAD for any ensemble containing the channel. This motivates us to identify conditions on the POVM effects and preparation states under which the associated Choi operator is positive definite.

For this, we examine the kernel of the Choi operator:
\begin{align} \label{eq-kernel_M&P}
    \text{ker}(\Gamma^{M,\mathcal{S}}) &= \text{ker}\left( \sum_{k=1}^n M_k^{T} \otimes \varsigma_k\right) = \bigcap_{k=1}^n \text{ker}\left( M_k^{T} \otimes \varsigma_k\right)
\end{align}

\begin{proposition}[\textbf{Positive definite measure-and-prepare Choi operators}]\label{prop-measure&prepare}
    For a measure-and-prepare channel with Choi operator $\Gamma^{M,\scal} = \sum_k M_k^T \otimes \varsigma_k$, each of the following is sufficient to satisfy $\Gamma^{M,\scal}>0$:
    \begin{enumerate}
        \item $\varsigma_k >0$ for all $\varsigma_k \in \scal$\label{item1-measure&prepare},
        \item $M_k >0$ for all $k$ and $\sum_k\varsigma_k >0$ \label{item2-measure&prepare},
        \item There exists a label $\alpha$ such that $\varsigma_{\alpha} >0$ and $M_{\alpha}>0$ \label{item3-measure&prepare}.
    \end{enumerate}
\end{proposition}
\begin{proof}
    For the proof, we use the facts pertaining to kernels of operators: 
    \begin{itemize}
        \item $\text{ker}(X_A +Y_A) = \text{ker}(X_A) \cap \text{ker}(Y_A)$ for $X_A,Y_A \ge 0$,
        \item $\text{ker}(X_A\otimes Y_B) = \text{ker}(X_A) \otimes \h_B$ for $Y_B>0$,
        \item $\text{ker}(X_A\otimes Y_B) = \h_A \otimes \text{ker}(Y_B)$ for $X_A >0$.
    \end{itemize}
    Proof of 1: If every $\varsigma_k >0$, then
    \begin{equation}
        \text{ker}(M_k^T \otimes \varsigma_k) = \text{ker}(M_k^T)\otimes \h_B,
    \end{equation}
    and therefore,
    \begin{align}
        \text{ker}(\Gamma^{M,\scal}) &= \bigcap_k \text{ker}(M_k^T) \otimes \h_B.
    \end{align}
    Since $\sum_k M_k = \mathbbm{1}_A$, we have $\bigcap_k\text{ker}(M_k^T) = \text{ker}(\mathbbm{1}_A) = \{0\}$. Hence $\text{ker}(\Gamma^{M,\scal}) = \{0\} \otimes \h_B = \{ 0\}$. \newline 
    \noindent Proof of 2: If every $M_k>0$, following the same arguments as before,
    \begin{align}
        \text{ker}(\Gamma^{M,\scal}) &= \h_A \otimes \text{ker}\left(\sum_k \varsigma_k\right).
    \end{align}
    This kernel is trivial only when $\sum_k\varsigma_k >0$. \newline     \noindent Proof of 3: Suppose there exists a label $\alpha$ such that $\varsigma_{\alpha}>0$ and $M_{\alpha}>0$. Since $M_{\alpha} > 0$ implies $M_{\alpha}^T > 0$, we have $M_{\alpha}^T \otimes \varsigma_{\alpha}>0$. Then,
    \begin{align}
        \Gamma^{M,\scal} &= M_{\alpha}^T \otimes \varsigma_{\alpha} + \sum_{k \neq \alpha} M_k^T \otimes \varsigma_{k}
    \end{align}
    is the sum of a positive definite operator and a positive semidefinite operator, and is therefore positive definite.
\end{proof}
Each condition in Proposition \ref{prop-measure&prepare} is sufficient to make the associated Choi operator positive definite. Together with Proposition \ref{prop-oneChoiFullRank}, this gives the following restrictions on strong- and weak-PAD.
\begin{coro}[\textbf{Obstruction to strong-PAD}]\label{coro:MP-strong}
    Any ensemble containing a measure-and-prepare channel that satisfies at least one of the criteria in Proposition \ref{prop-measure&prepare} cannot be strongly-PAD.
\end{coro}

\begin{coro}[\textbf{Obstruction to weak-PAD}]\label{coro:MP-weak}
    An ensemble in which all candidates except at most one are measure-and-prepare channels satisfying at least one of the criteria in Proposition \ref{prop-measure&prepare} cannot be weakly-PAD.
\end{coro}

\begin{example}[\textbf{Preparation set of depolarised states}] If the preparation set contains depolarised states of the form,
\begin{equation}
    \tilde{\varsigma}_k = (1-p_k)\varsigma_k + p_k \frac{\id_B}{d_B}; \quad p_k \in (0,1] \text{ for all }k,
\end{equation}
        then $\tilde{\varsigma}_k >0$, and therefore Statement \ref{item1-measure&prepare} of Proposition \ref{prop-measure&prepare} immediately gives that the Choi operator of the associated measure-and-prepare channel is positive definite. Physically, this means that any measure-and-prepare channel in which each and every preparation state contains a nonzero white-noise component, $p_k > 0$ for all $k$, obstructs strong-PAD of every ensemble it is a candidate of.
\end{example}

Another instance of Proposition \ref{prop-measure&prepare} arises for quantum-to-classical channels, where the preparation states are fixed by the outcome labels of the corresponding POVM.

\begin{example}[\textbf{Quantum-to-classical channels}] \label{example:quantum-to-classical}
    The quantum-to-classical measure-and-prepare channel corresponding to any
    $n-$outcome POVM $M$ has the classical states $\mathcal{S} = \{ \dyad{k}\}_{k=1}^n$, where $\{ \ket{k}\}_{k=1}^n$ is the computational basis for $\mathbb{C}^n$. We can note that $\text{ker}(\sum_{k=1}^n \dyad{k}) = \text{ker}(\id_B) = \{ 0\}$, where the output system $B$ is the $n$-dimensional classical register. Consequently, the Choi operator $\Gamma^{M,\mathcal{S}}$ of any such channel, associated with a POVM whose effects are positive definite, is positive definite (Also see from Proposition \ref{prop-measure&prepare} Statement \ref{item2-measure&prepare}). An important class of POVMs with positive definite effects is that of noisy versions of any given one $N$,
    \begin{equation}
        \tilde{N}_k = (1-p) N_k + \frac{p}{n}\id; \quad p \in (0,1).
    \end{equation}
\end{example}

For qubit measurements, the above restriction can be illustrated more explicitly by considering binary POVMs and their sharpness parameter.

\begin{example}[\textbf{Unsharp binary qubit POVMs}] 
The two effects of any arbitrary unbiased binary POVM \cite{busch2016quantum} admit the following Bloch form:
\begin{align}
    M_{\pm} = \frac{1}{2}(\id \pm \eta \boldsymbol{n}\cdot \boldsymbol{P}) ~:~ 0 \le \eta \le 1, \norm{\boldsymbol{n}}=1,
\end{align}
where $\boldsymbol{P} = (X,Y,Z)$ and $\boldsymbol{n} = (n_X, n_Y, n_Z)$. $\eta $ is the sharpness parameter or visibility; The POVM becomes sharp or projective for $\eta =1$ and completely unsharp for $\eta = 0$. The eigenvalues of $M_{\pm}$ are $\{ \frac{1 \pm \eta}{2}, \frac{1 \mp \eta}{2} \}$. Therefore, the POVM is strictly positive for $\eta \neq 1$. Consequently, the quantum-to-classical channel associated with a binary qubit POVM with non-unit sharpness has a positive definite Choi operator, and no ensemble containing it admits strong-PAD.
\end{example}

The preceding results identify several instances where the candidature of measure-and-prepare channels obstructs perfect anti-distinguishability. We now consider a particular family of quantum-to-classical channels arising from unlabeled POVMs. In this case, the possible associations between measurement effects and outcome labels define an ensemble of channels, whose anti-distinguishability can be studied using the same structural conditions.

\subsection{Unlabeled POVMs}
An unlabeled POVM is a partial description of a quantum state measurement, for which the associations between outcome labels and POVM effects are unknown or undercharacterised. Sudarsanan Ragini et al. \cite{Nidhin2024SingleShotLabeling, Arash2024MultipleShotLabeling} formulate an unlabeled $n$-outcome POVM as the $n!$ possible permutation (quantum-to-classical) channels, under the assumption that all involved effects furnished by a reference POVM $M$ are non-identical. These channels are defined by their action as
\begin{align}
    \mathcal{M}_{\sigma}(\varrho) := \sum_{i=1}^n \text{Tr}[\varrho M_i] \dyad{x_{\sigma(i)}},
\end{align}
with Choi operators
\begin{align}
    \Gamma_{\sigma} = \sum_{i=1}^n M_i^T \otimes \dyad{x_{\sigma(i)}}.
\end{align}
The authors then formulate ``labeling'' of the unlabeled POVM as a distinguishability problem of the above ensemble of channels, after assigning equal priors to each candidate hypothesis. 

Here, we study the complement antidistinguishability aspect pertaining to  said ensembles. So, instead of labeling the POVM through distinguishability, antidistinguishability should exclude possible label-effect associations. We find a necessary and sufficient condition for strong-PAD for the ensemble $\xcal_M$ associated to any $M$,
\begin{align}
    \xcal_M = \{ \mathcal{M}_{\sigma}\}_{\sigma \in S_n},
\end{align}
where $S_n$ is the permutation group over the $n$ outcome labels.
\begin{proposition}[\textbf{Exact conditions for strong-PAD}]
    An ensemble $\xcal_{M}$ admits strong-PAD if and only if there exists at least one outcome label $a$ such that $M_a$ is rank-deficient.
\end{proposition}
\begin{proof}
Assume each and every effect of $M$ is positive definite. Then, from Proposition \ref{prop-measure&prepare} Statement \ref{item2-measure&prepare}  or  Example \ref{example:quantum-to-classical}, $\Gamma_{\sigma}$ is positive definite for all $\sigma \in S_n$ and thus $\xcal_M$ does not even admit weak-PAD. From this, we can conclude that necessarily at least one effect should be rank-deficient. 

Let us now show sufficiency. Suppose there exists an effect $M_a$ which is rank-deficient, that is 
\begin{align}
    \text{ker}(M_a) \neq \{ 0\}.
\end{align}
Choose a state vector from this kernel, $\ket{\varphi} \in \text{ker}(M_a)$ such that it satisfies $M_a \ket{\varphi} =0$. Since the Choi operators $\Gamma_{\sigma}$ have $M_a^T$, define the PPOVM normalisation state
\begin{align}
    \xi_A = \dyad{\Bar{\varphi}},
\end{align}
where the complex conjugation is with respect to the same basis from which the transpose in $M_a^T$ is also defined. Thus, we have
\begin{align}
    M_a^T \ket{\bar{\varphi}} = 0
\end{align}
and hence
\begin{align}
    M_a^T \xi_A = 0 \implies \text{Tr}[M_a^T \xi_A] = 0.
\end{align}
For every permutation $\sigma \in S_n$, define
\begin{align}
    T_{\sigma} := \frac{1}{(n-1)!} \xi_A \otimes \dyad{x_{\sigma(a)}}.
\end{align}
By construction, \(T_\sigma\geq 0\). To verify the PPOVM normalisation, we regroup the sum over the permutation group \(S_n\) according to the output label \(\sigma(a)\). For each fixed \(y\in\{1,\ldots,n\}\), there are exactly \((n-1)!\) permutations satisfying \(\sigma(a)=y\), since the remaining \(n-1\) labels can be permuted arbitrarily. Thus, the sum over permutations can be replaced by a sum over output labels, with each label contributing a multiplicity factor of \((n-1)!\). Explicitly,
\begin{align}
    \sum_{\sigma \in S_n} T_{\sigma} &= \sum_{\sigma \in S_n}\frac{1}{(n-1)!} \xi_A \otimes \dyad{x_{\sigma(a)}} \\
    &= \frac{1}{(n-1)!} \xi_A  \otimes \left( \sum_{\sigma \in S_n}\dyad{x_{\sigma(a)}}\right) \\
    &= \frac{1}{(n-1)!} \xi_A \otimes \sum_{y=1}^n \sum_{\substack{\sigma \in S_n \\ \sigma(a) = y}} \dyad{x_y}\\
    &= \frac{1}{(n-1)!} \xi_A \otimes \left( (n-1)! \sum_{y=1}^n \dyad{x_y} \right) \\
    &= \xi_A \otimes \sum_{y=1}^n \dyad{x_y} ~= ~\xi_A \otimes \id_B.
\end{align}
For any permutation $\sigma$, the Choi operator $\Gamma_{\sigma}$ is block diagonal with respect to the output basis,
\begin{align}
    \Gamma_{\sigma} = \bigoplus_{y=1}^n M^T_{\sigma^{-1}(y)}.
\end{align}
Therefore, for a fixed permutation $\sigma$, in $\Gamma_{\sigma}$, the block associated with output $x_{\sigma(a)}$ is exactly $M_a^T$. Hence,
\begin{align}
    \text{Tr}[\Gamma_{\sigma}T_{\sigma}] &= \frac{1}{(n-1)!} \text{Tr}[M^T_a \xi_A] \\
    &= 0.
\end{align}
Therefore, every PPOVM outcome $\neg \sigma$ conclusively excludes the candidate hypothesis $\mathcal{M}_{\sigma}$, thus satisfying weak-PAD. Now, it remains to verify satisfaction of strong-PAD. Fix $\sigma$, and consider
\begin{align}
    \sum_{\tau \in S_n} \text{Tr}[\Gamma_{\tau} T_{\sigma}]. 
\end{align}
Define $y = \sigma(a)$. For a given $\tau$, the block of $\Gamma_{\tau}$ corresponding to output $x_y$ is $M^T_{\tau^{-1}(y)}$. Thus,
\begin{align}
    \text{Tr}[\Gamma_{\tau} T_{\sigma}] &= \frac{1}{(n-1)!} \text{Tr}\left[ M^T_{\tau^{-1}(y)} \xi_A\right].
\end{align}
We again regroup the sum over $S_n$, now according to the value of $\tau^{-1}(y)$. For each fixed index $i$, exactly $(n-1)!$ permutations satisfy $\tau^{-1}(y) =i$, equivalently $\tau(i) = y$. Consequently, every POVM effect $M_i$ contributes with the same multiplicity, furnishing
\begin{align}
    \sum_{\tau \in S_n} \text{Tr}[\Gamma_{\tau} T_{\sigma}] &= \sum_{\tau \in S_n}\frac{1}{(n-1)!} \text{Tr}\left[ M^T_{\tau^{-1}(y)} \xi_A\right] \\
    &= (n-1)! \frac{1}{(n-1)!} \sum_{i=1}^n \text{Tr}[M_i^T \xi_A] \\
    &= \text{Tr}\left[ \left( \sum_{i=1}^n M_i^T\right)\xi_A\right] \\
    &= \text{Tr}[\id_A \xi_A] \\
    &= 1.
\end{align}
Moreover, 
\begin{align}
    \sum_{\tau \in S_n} \text{Tr}[\Gamma_{\tau} T_{\sigma}] > 0 ~~\forall \sigma.
\end{align}
Thus, every exclusion outcome occurs with non-zero probability for at least one candidate channel, and $\xcal_M$ admits strong-PAD with the constructed PPOVM.
\end{proof}

Thus, for the permutation-channel ensemble associated with an unlabeled POVM, the existence of a rank-deficient effect is both necessary and sufficient for strong-PAD. In particular, the condition can be verified directly from the effects of the reference POVM, without separately examining all the associated permutation channels.

\begin{example}[\textbf{Rank-one POVMs}] \label{example:unlabeled-POVM-instances}
Let $M$ be an $n$-outcome POVM with non-identical rank-one effects $M_i=w_i\dyad{\psi_i}$, where $w_i>0$. Every effect is rank-deficient, so the associated permutation-channel ensemble $\xcal_M$ admits strong-PAD. More explicitly, fix an effect $M_a$ and a normalised vector $\ket{\varphi}$ orthogonal to $\ket{\psi_a}$. The unassisted probe $\dyad{\varphi}$ gives the PPOVM,
\begin{align}
    T_{\sigma}=\frac{1}{(n-1)!}\dyad{\bar{\varphi}}_A\otimes\dyad{x_{\sigma(a)}}_B.
\end{align}
As in the preceding proof, these effects have normalisation $\dyad{\bar{\varphi}}_A\otimes\id_B$ and satisfy,
\begin{align}
    \text{Tr}[\Gamma_{\sigma}T_{\sigma}]&=0, \\
    \sum_{\tau\in S_n}\text{Tr}[\Gamma_{\tau}T_{\sigma}]&=1\qquad\forall\ \sigma.
\end{align}
Operationally, an observed output label $x_y$ excludes every permutation assigning effect $M_a$ to that label. Uniformly choosing one of these $(n-1)!$ permutations gives the above PPOVM.

(\textbf{von Neumann measurements})\textbf{.}
 For the qudit POVM $M_i=\dyad{i}$, $i=1,\dots,d$, choose the probe $\dyad{b}$ with $b\neq a$. Effect $M_a$ then has zero probability, and the construction above strongly anti-distinguishes the $d!$ permutation channels. For $d=2$, there are only two permutations, and this coincides with perfect distinguishability. For $d>2$, the probe identifies the label assigned to effect $M_b$, while the assignments of the remaining effects are undetermined. Nevertheless, every observed label still excludes any permutation that assigns $M_a$ to it.

(\textbf{Qubit trine POVM})\textbf{.} With $\omega=\exp(2\pi i/3)$, take,
\begin{align}
    \ket{\psi_k}&=\frac{1}{\sqrt{2}}\big(\ket{0}+\omega^{k-1}\ket{1}\big), \\
    M_k&=\frac{2}{3}\dyad{\psi_k},\qquad k=1,2,3.
\end{align}
These effects sum to $\id_A$. Choosing $a=1$ and $\ket{\varphi}=(\ket{0}-\ket{1})/\sqrt{2}$ gives outcome probabilities $0,\frac{1}{2},\frac{1}{2}$ for the reference POVM. Thus, the ensemble of six permutation channels admits strong-PAD with,
\begin{align}
    T_{\sigma}=\frac{1}{2}\dyad{\varphi}_A\otimes\dyad{x_{\sigma(1)}}_B,\qquad\sigma\in S_3.
\end{align}
Each observed output label excludes the two permutations assigning $M_1$ to that label.

(\textbf{Qubit symmetric informationally complete (SIC) POVM})\textbf{.}  An explicit four-outcome instance is given by $M_i=\frac{1}{2}\dyad{\psi_i}$, with,
\begin{align}
    \ket{\psi_1}&=\ket{0}, \\
    \ket{\psi_{k+2}}&=\frac{1}{\sqrt{3}}\ket{0}+\sqrt{\frac{2}{3}}\omega^k\ket{1},\quad k=0,1,2.
\end{align}
The effects sum to $\id_A$. Due to the said symmetry, state vectors satisfy $\abs{\braket{\psi_i}{\psi_j}}^2=\frac{1}{3}$ for $i\neq j$. With $a=1$ and probe $\dyad{1}$, the reference outcome probabilities are $0,\frac{1}{3},\frac{1}{3},\frac{1}{3}$. Consequently, the ensemble of $24$ permutation channels admits strong-PAD with,
\begin{align}
    T_{\sigma}=\frac{1}{6}\dyad{1}_A\otimes\dyad{x_{\sigma(1)}}_B,\qquad\sigma\in S_4.
\end{align}
Here, each observed output label excludes six permutations. These examples concern the exclusion of the possible label-effect associations, as represented by the permutation channels, with the probes and effects above explicitly realising this task.
\end{example}
\section{State-space contractions}\label{sec:state-space-contractions}
State-space contraction channels model a wide class of quantum noise where an input state is probabilistically replaced by another fixed ``noise state''. As such, having a fixed noise state $\nu$ and  knowledge that it occurs with a non-zero probability $q$, we have the following definition,
\begin{definition}[\textbf{State-space contractions}]
    With isomorphic input and output systems $A$ and $B$, given $(q,\nu)$ with $0<q\le1$, associated probabilistic state-space contraction $\ccal^{(q,\nu)}_{A \rightarrow B}$ is defined through its action as,
    \begin{equation}
        \ccal^{(q,\nu)}_{A \rightarrow B}(\varrho_A) = (1-q)\varrho_B + q \text{Tr}(\varrho)\nu_B.
    \end{equation}
\end{definition}
The Choi operator of the above channel is given by,
\begin{equation}
   \Gamma^{{(q,\nu)}}_{AB}= (1-q)\Gamma_{AB} + q (\id_A \otimes\nu_B).
\end{equation}
When $q=1$, any arbitrary input state is completely replaced by $\nu$. Such a  channel is also referred to as a complete state-space contraction \cite{heinosaari2011mathematical},  a replacer channel or a trash-and-prepare channel.

We first consider when the inclusion of a state-space contraction in an ensemble is sufficient to preclude strong-PAD. The following proposition identifies such a condition in terms of the noise state of the channel.

\begin{proposition}[\textbf{Obstruction due to candidature}]
    An ensemble having a state-space contraction  $\ccal^{(q,\nu)}$ as a candidate, with a positive definite  noise state $\nu >0$, does not admit strong-PAD for any $q \in (0,1]$.
\end{proposition}
\begin{proof}
   The kernel of the Choi operator, $\text{ker}(\Gamma^{{(q,\nu)}}) = \text{ker}(\Gamma + \id \otimes \nu) = \text{ker}(\Gamma) \cap \text{ker}(\id \otimes \nu)$.  But, we have $\text{ker}(\id \otimes \nu) = \{ 0\}$ and, consequently, $\text{ker}(\Gamma^{{(q,\nu)}}) = \{ 0\}$, thus $\Gamma^{{(q,\nu)}} >0$. This translates to inadmissibility of strong-PAD due to Proposition \ref{prop-oneChoiFullRank}.
\end{proof}

\begin{example}[\textbf{Depolarisers}]
An important family of state-space contractions is that of depolarisers, where the noise state is $\nu = \frac{\id}{d}$. The depolariser that completely replaces the input state with the maximally mixed ($q=1$), is the complete depolariser. Inclusion of any of these channels sabotages strong-PAD for any ensemble.
\end{example}

The preceding obstruction concerns strong-PAD and follows from the positive definiteness of the Choi operator of a candidate channel. We now consider a different restriction that applies to ensembles consisting solely of state-space contractions. In particular, when each candidate contains a nonzero contribution from the identity channel, the common mixing result of Proposition \ref{prop:mix-a-fixed-Choi} allows us to rule out even weak-PAD.

\begin{proposition}[\textbf{Obstruction due to common noise}]
    An ensemble of only state-space contractions $\{\Gamma^{{(q_j,\nu_j)}}\}_{j=1}^N$, with $q_j <1$ for all $j$, does not admit weak-PAD.
\end{proposition}
\begin{proof}
    The proof is a consequence of Proposition \ref{prop:mix-a-fixed-Choi} since each and every candidate $\Gamma^{{(q_j,\nu_j)}}$ can be seen as a physical mixture containing the same Choi operator $\Gamma$, that of the identity channel.
\end{proof}

Having established that ensembles of state-space contractions with a nonzero identity contribution do not admit weak-PAD, we now consider ensembles where every candidate is a complete state-space contraction. These are precisely the replacer channels discussed above. For such ensembles, perfect anti-distinguishability reduces to that of the associated replacement states.

The Choi operator of a replacer channel with replacement state $\nu$ is
\begin{align}
    \Gamma_{AB}^{(\nu)} = \id_A \otimes \nu_B.
\end{align}
\begin{proposition}\label{prop:replacer-state<->channel} An ensemble of only replacer channels $\{\Gamma^{(\nu_j)}\}_{j}$ is weak/strong-PAD if and only if the associated ensemble of replacement states $\{\nu_j\}_j$ is respectively weak/strong-PAD.
\end{proposition}

\begin{proof}
Suppose that a PPOVM $T$, with normalisation $\xi_A\otimes\id_B$, weakly PAD the replacer-channel ensemble. Define operators on the output system by $M_j:=\text{Tr}_A[T_j]$. Since partial trace preserves positivity, $M_j\geq 0$ for every $j$. Moreover,
\begin{align}
    \sum_j M_j &= \text{Tr}_A\left[\sum_jT_j\right] \\
    &= \text{Tr}_A[\xi_A\otimes\id_B]=\id_B,
\end{align}
so $M$ is a POVM. Using $\Gamma^{(\nu_k)}=\id_A\otimes\nu_k$, we obtain, for every $j,k$,
\begin{align}
    \text{Tr}[\Gamma^{(\nu_k)}T_j]
    &= \text{Tr}[(\id_A\otimes\nu_k)T_j] \\
    &= \text{Tr}[\nu_kM_j].
\end{align}
In particular, the weak-PAD conditions for $T$ give $\text{Tr}[\nu_jM_j]=0$ for all $j$, so $M$ weakly anti-distinguishes the replacement states. If $T$ satisfies strong-PAD as well, then,
\begin{align}
    \sum_k\text{Tr}[\nu_kM_j]
    =\sum_k\text{Tr}[\Gamma^{(\nu_k)}T_j]>0
\end{align}
for every $j$. Thus, $M$ also satisfies the strong-PAD conditions for the state ensemble.

Conversely, suppose that a POVM $M$ weakly or strongly anti-distinguishes the replacement states. Choose any normalisation state $\xi_A$ and define $T_j:=\xi_A\otimes M_j$. These operators are positive and satisfy,
\begin{align}
    \sum_jT_j=\xi_A\otimes\sum_jM_j=\xi_A\otimes\id_B.
\end{align}
Hence, they form a PPOVM. For every $j,k$,
\begin{align}
    \text{Tr}[\Gamma^{(\nu_k)}T_j]
    &=\text{Tr}[\xi_A]\text{Tr}[\nu_kM_j] \\
    &=\text{Tr}[\nu_kM_j].
\end{align}
Therefore, the vanishing probabilities required for weak-PAD and the nonzero sums required for strong-PAD are preserved. The replacer-channel ensemble consequently admits the same weak or strong form of perfect anti-distinguishability as the replacement-state ensemble, establishing both implications.
\end{proof}

The preceding analyses concerned particular families of channels. We now consider ensembles characterised by the pairwise commutativity of their Choi operators. This allows the necessary condition involving the common Choi support to be expressed in terms of the product of the candidate Choi operators.

\section{Choi-commuting ensembles}\label{sec:choi-commuting}
A tractable, yet physically non-trivial, class of channel ensembles is obtained when all of the candidate Choi operators pairwise commute. Such a restriction is not imposed merely for algebraic tractability. Several familiar families of quantum channels naturally create such ensembles; Some canonical examples being Pauli channels, and more generally channels whose Choi operators are diagonal with respect to a Hilbert-Schmidt orthogonal error basis, such as generalized Weyl-Heisenberg channels \cite{FukudaHolevo2005, DattaFukudaHolevo2006}. Other instances can be found in classical stochastic channels, suitable families of measure-and-prepare channels, erasure channels, etc. Thus, this ``Choi commutativity’’ concerns a broad collection of quantum processes, ranging from classical noise models to genuinely quantum random-unitaries \cite{AudenaertScheel2008}.  To proceed, let us formally define this notion of Choi commutativity:
\begin{definition}[\textbf{Choi-commuting ensembles}]
    An ensemble $\xcal$ is ``Choi-commuting" if all candidate Choi operators pairwise commute with each other; i.e.,
    \begin{align}
        [\Gamma_x, \Gamma_y] = 0 ~~~\forall \Gamma_x,\Gamma_y \in \xcal.
    \end{align}
\end{definition}
Since a Choi-commuting ensemble pairwise commutes, each and every Choi operator admit simultaneous diagonalisation. That is, spectral decomposition of these positive operators identifies a common orthonormal eigenbasis $\{\ket{r_j}\}_{j=1}^{d_A d_B}$ of $\h_A \otimes \h_B$ such that for all $x$,
    \begin{align}
        \Gamma_x = \sum_{i=1}^{d_A d_B} \lambda_i^{(x)} \dyad{r_i} \text{ with } \lambda_i^{(x)} \ge 0 \ \forall i.
    \end{align}

Simultaneous diagonalisation allows us to relate the common Choi support to the support of the product of the candidate Choi operators. The following lemma establishes this relation, which we then use to express the necessary condition from Theorem \ref{theorem-PAD_necessary} in terms of this product.

  \begin{lemma}
      For a Choi-commuting ensemble, support of product of all Choi operators coincides with the common Choi support. 
  \end{lemma}
  \begin{proof}
      For each candidate label $x$, define the set of labels
      \begin{align}
          \Omega^{(x)} := \{ j\hspace{0.05cm}:\hspace{0.05cm} \lambda_j^{(x)} >0\}.
      \end{align}
      Then, the common Choi support of the ensemble is 
      \begin{align}
          S_{\xcal} &= \bigcap_{x=1}^n \text{supp}(\Gamma_x) \\
          &= \text{span}\{ \ket{r_j}\hspace{0.05cm}:\hspace{0.05cm} j \in \bigcap_{x}\Omega^{(x)}\}.
      \end{align}
Now define the product of all the candidate Choi operators,
\begin{align}
    P_{\xcal} := \prod_{x=1}^n \Gamma_x.
\end{align}
Since all the candidate Choi operators are diagonal in the common orthonormal eigenbasis $\{\ket{r_j}\}$, this product reads
\begin{align}
    P_{\xcal} &= \sum_{j} \mu_j\ketbra{r_j}{r_j} \text{ with,} \\
        \mu_j &:= \left( \prod_{x=1}^n \lambda_j^{(x)} \right).
\end{align}
Since every $\lambda_j^{(x)} \ge 0$, if the eigenvalue corresponding to $j^{\mathrm{th}}$ eigenstate of the product $P_{\xcal}$ is positive then $\lambda_j^{(x)} >0 ~~\forall x$. Equivalently, 
\begin{align}
    j \in \bigcap_{x} \Omega^{(x)}.
\end{align}
Hence the support of the product,
\begin{align}
    \text{supp}(P_{\xcal}) &= \text{span}\{\ket{r_j}\hspace{0.05cm}:\hspace{0.05cm} \mu_j>0\} \\
    &= \text{span}\left\{\ket{r_j}\hspace{0.05cm}:\hspace{0.05cm}j \in\bigcap_{x} \Omega^{(x)}\right\},
\end{align}
coincides with $S_{\xcal}$.
  \end{proof}
  Moreover, we have trivial common Choi support $S_{\xcal} = \{ 0\}$ if and only if there is no common eigenvector having strictly positive eigenvalue for every candidate Choi operator, that is when
  \begin{align}
      \bigcap_{x} \Omega^{(x)} = \varnothing.
  \end{align}

Equipped with the above lemma, we can express the necessary condition on the common Choi support in terms of the product of the candidate Choi operators and the PPOVM normalisation. This gives the following condition for weak-PAD of Choi-commuting ensembles.

  \begin{proposition}[\textbf{A necessity for Choi-commuting ensembles}]\label{prop-CommutingCHoiNecessary}
      Let $\xcal$ be  Choi-commuting. If a PPOVM with normalisation $\xi_A \otimes \id_B$ weakly-PAD $\xcal$ then necessarily
      \begin{align}
          P_{\xcal} (\xi_A \otimes \id_B)= 0.
      \end{align}
  \end{proposition}
  \begin{proof}
      Let $T$ be a PPOVM, with normalisation $(\xi_A \otimes \id_B)$, that weakly-PAD $\xcal$. Since $T_x,\Gamma_x \ge 0$ for all $x$, the weakly-PAD conditions translate to the operator equalities,
      \begin{align}
          \Gamma_x T_x = 0~~\forall x.
      \end{align}
      Now, for a fixed $x$,
      \begin{align}
          P_{\xcal} T_x &= \left(\prod_{y \neq x} \Gamma_y\right) \underbrace{\Gamma_x T_x}_{= 0} = 0.
      \end{align}
      Since this holds for every $x$, summing over all PPOVM outcomes and invoking associated normalisation $\sum_x T_x = \xi_A \otimes \id_B$,
      \begin{align}
          0 &= \sum_x P_{\xcal} T_x \\
          &= P_{\xcal} \left( \sum_x T_x\right) \\
          &= P_{\xcal} (\xi_A \otimes \id_B).
      \end{align}
  \end{proof}

The condition in Proposition \ref{prop-CommutingCHoiNecessary} need not be sufficient for weak-PAD with a given PPOVM normalisation. We illustrate this with two qubit unitary channels.

  \begin{example}
      Consider the two qubit unitary channels generated by  $\id $ and $Z$ (the identity channel and the phase-flip respectively). Respective Choi operators are,
      \begin{align}
          \Gamma_{\id} = \dyad{\Gamma} \text{ and, } \Gamma_{Z} = \dyad{Z},
      \end{align}
      with $\ket{Z} := (\id \otimes Z) \ket{\Gamma}$. Consequently we have, $\braket{\Gamma}{Z} = \text{Tr}[Z] = 0.$ Therefore, the product of the two Choi operators,
      \begin{align}
          P_{\xcal} = \Gamma_{\id} \Gamma_Z = 0.
      \end{align}
      Hence the condition, 
      \begin{align}
          P_{\xcal} (\xi \otimes \id) = 0
      \end{align}
     is satisfied for any normalisation state $\xi$. However, consider the unassisted normalisation $\dyad{0}\otimes \id$, which operationally corresponds to feeding the state $\ket{0}$ into the channels as the probe. Both channels then output the same state $\dyad{0}$. For two channels, this means that, with this probe, they can neither be perfectly distinguished nor, equivalently, perfectly anti-distinguished.
  \end{example}

The preceding example shows that the condition in Proposition \ref{prop-CommutingCHoiNecessary} is not sufficient for arbitrary PPOVM normalisations. However, when the normalisation also commutes with every candidate Choi operator, the same condition becomes sufficient, as established next.

  \begin{theorem}[\textbf{Exact condition for Choi-commuting ensembles}]\label{theorem:Choi-commuting-weak-PAD}
      Let $\xcal$ be Choi commuting, and let  $\xi_A \otimes \id_B$ be a valid PPOVM normalisation that pairwise commutes with each and every candidate Choi operator. Then, $\xcal$ admits weak-PAD by a PPOVM with normalisation $\xi_A \otimes \id_B$ if and only if
      \begin{align}
          P_{\xcal} (\xi_A \otimes \id_B) = 0.
      \end{align}
  \end{theorem}
  \begin{proof}
  The forward implication is already proved by Proposition \ref{prop-CommutingCHoiNecessary} for any arbitrary normalisation.  For the converse, 
  suppose 
  \begin{align}
      P_{\xcal} (\xi_A \otimes \id_B) = 0.
  \end{align}
  Since $[\Gamma_x,\Gamma_y] = 0 ~\forall \Gamma_x,\Gamma_y \in \xcal$ and with the commutativity assumption 
  \begin{align}
      [\xi_A\otimes \id_B, \Gamma_x ] = 0~~\forall \Gamma_x \in \xcal,
  \end{align}
  all of the Choi operators, along with the normalisation, are simultaneously diagonalisable; The normalisation can then be represented as
  \begin{align}
      \xi_A \otimes \id_B = \sum_{j} \nu_j \ketbra{r_j}{r_j} ~~: \nu_j \ge 0.
  \end{align}
  Right multiplying the product $P_{\xcal}$ with the normalisation,
  \begin{align}
      P_{\xcal} (\xi_A \otimes \id_B) = \sum_j \nu_j \left( \prod_{x} \lambda_{j}^{(x)}\right) \dyad{r_j}.
  \end{align}
          Since this operator vanishes,
          \begin{align}
                            \nu_j \left( \prod_{x} \lambda_{j}^{(x)}\right)  = 0 ~~\forall j.
          \end{align}
          Now, consider a specific $j$ with positive $\nu_j >0$. This implies that
          \begin{align}
               \prod_{x} \lambda_{j}^{(x)} = 0.
          \end{align}
          Since all of the factors appearing above are non-negative, there must exist at least one $x$ for which $\lambda_j^{(x)}=0$. For each $j$, collect such labels $x$ and define label sets
          \begin{align}
          \mathcal{A}_j := \{x \hspace{0.05cm}:\hspace{0.05cm} \lambda_j^{(x)}=0 \}.
          \end{align}
          Therefore, for every $j$ supported by the normalisation, we have
          \begin{align}
              \nu_j >0 \implies \mathcal{A}_j \neq \varnothing.
          \end{align}
          We construct the PPOVM effects:
          \begin{align}
              T_x := \sum_{\substack{j:\nu_j>0 \\ x \in \mathcal{A}_j}} \frac{\nu_j}{\abs{\mathcal{A}_j}} \dyad{r_j}.
          \end{align}
          This construction can be understood as distributing the spectral weight of the normalisation among the candidate channels. For each eigenvector \(|r_j\rangle\) supported by the normalisation, the set $\mathcal{A}_j$ identifies precisely those candidates whose Choi operators vanish on that eigenvector. We distribute the corresponding weight \(\nu_j\) uniformly among these candidates. This guarantees that each assigned contribution is orthogonal to the Choi operator of the candidate it excludes, while the total weight is preserved under summation over all exclusion outcomes.
          
          By construction, $T_x \ge 0$ for all $x$ and moreover we achieve the desired normalisation,
          \begin{align}
              \sum_x T_x &=  \sum_x \sum_{\substack{j:\nu_j>0 \\ x \in \mathcal{A}_j}} \frac{\nu_j}{\abs{\mathcal{A}_j}} \dyad{r_j} \\
              &= \sum_{j: \nu_j >0}\sum_{x \in \mathcal{A}_j}\frac{\nu_j}{\abs{\mathcal{A}_j}} \dyad{r_j} \\
              &= \sum_{j:\nu_j>0 }\nu_j \dyad{r_j}\\
              &= \xi_A \otimes \id_B.
          \end{align}
          Now, 
          \begin{align}
              \Gamma_x T_x = \sum_{\substack{j:\nu_j>0 \\ x \in \mathcal{A}_j}} \lambda_j^{(x)}\frac{\nu_j}{\abs{\mathcal{A}_j}} \dyad{r_j}.
          \end{align}
          Since $x \in \mathcal{A}_j$ implies $\lambda_j^{(x)} =0$, we equivalently have the following,
          \begin{align}
              \Gamma_x T_x &= 0 ~~\forall x, \\
              \text{Tr}[\Gamma_x T_x] &=0 ~~ \forall x,
          \end{align}
          which are the weak-PAD conditions and therefore $\xcal$ is weak-PAD with a PPOVM with normalisation $\xi_A \otimes \id_B$.
  \end{proof}

An important instance of Choi-commuting ensembles is provided by Pauli channels. Since their Choi operators are diagonal in a common basis, the preceding conditions can be examined directly in terms of the associated Pauli probabilities.

\begin{example}[\textbf{Qubit Pauli channels}]
Let $\xcal = \{\pcal^{\Vec{p}^{(x)}}\}_{x=1}^n$ be an ensemble of qubit Pauli channels. With $\mathcal{B} = \{\sigma_0,\sigma_X,\sigma_Y,\sigma_Z\} = \{\id,X,Y,Z\}$, their action is given by,
\begin{align}
    \pcal^{\Vec{p}^{(x)}}(\varrho) = \sum_{\alpha\in\{0,X,Y,Z\}} p_{\alpha}^{(x)}\sigma_{\alpha}\varrho\sigma_{\alpha}^{\dagger},
\end{align}
where $p_{\alpha}^{(x)}\geq 0$ and $\sum_{\alpha}p_{\alpha}^{(x)}=1$ for every $x$. The normalised Bell vectors $\ket{r_{\alpha}}$ of Eq.(\ref{eq:bell-vectors}) form a common orthonormal eigenbasis for the candidate Choi operators. In the Choi convention used here,
\begin{align}
    \Gamma_x &= 2\sum_{\alpha}p_{\alpha}^{(x)}\dyad{r_{\alpha}}, \\
    P_{\xcal} &= 2^n\sum_{\alpha}\left(\prod_{x=1}^n p_{\alpha}^{(x)}\right)\dyad{r_{\alpha}}.
\end{align}
The maximally-assisted normalisation $\frac{\id_A}{2}\otimes\id_B$ commutes with every $\Gamma_x$. Consequently,  Theorem \ref{theorem:Choi-commuting-weak-PAD} gives that $\xcal$ admits weak-PAD with this normalisation if and only if,
\begin{align}\label{eq:pauli-weak-PAD}
    \prod_{x=1}^n p_{\alpha}^{(x)}=0 \qquad \forall\ \alpha\in\{0,X,Y,Z\}.
\end{align}
That is, for each Pauli label, at least one candidate assigns zero probability to the corresponding unitary. In fact, this condition is necessary for any PPOVM: if $p_{\alpha}^{(x)}>0$ for every $x$ at some fixed $\alpha$, then every candidate contains a nonzero contribution from the same Pauli unitary channel. Proposition \ref{prop:mix-a-fixed-Choi} then precludes even weak-PAD, irrespective of the normalisation.

When Eq.(\ref{eq:pauli-weak-PAD}) holds, an explicit PPOVM follows from the construction in the  proof from above. For each $\alpha$, define the nonempty set $\mathcal{A}_{\alpha}:=\{x:p_{\alpha}^{(x)}=0\}$ and choose,
\begin{align}
    T_x := \frac{1}{2}\sum_{\alpha:\,x\in\mathcal{A}_{\alpha}}\frac{\dyad{r_{\alpha}}}{\abs{\mathcal{A}_{\alpha}}}.
\end{align}
Then $\sum_xT_x=\frac{\id_A}{2}\otimes\id_B$ and $\text{Tr}[\Gamma_xT_x]=0$ for all $x$. Operationally, one uses a maximally entangled probe and measures the output in the Bell basis. An observed label $\alpha$ is assigned uniformly to one of the candidates in $\mathcal{A}_{\alpha}$ for exclusion, since that outcome has zero probability for each such candidate.

For a concrete instance, consider three channels with probability vectors in the order $(0,X,Y,Z)$,
\begin{align}
    \Vec{p}^{(1)} &= \left(\frac{1}{2},\frac{1}{2},0,0\right), \\
    \Vec{p}^{(2)} &= \left(0,\frac{1}{2},\frac{1}{2},0\right), \\
    \Vec{p}^{(3)} &= \left(\frac{1}{2},0,\frac{1}{2},0\right).
\end{align}
The above construction yields,
\begin{align}
    T_1 &= \frac{1}{2}\dyad{r_Y}+\frac{1}{6}\dyad{r_Z}, \\
    T_2 &= \frac{1}{2}\dyad{r_0}+\frac{1}{6}\dyad{r_Z}, \\
    T_3 &= \frac{1}{2}\dyad{r_X}+\frac{1}{6}\dyad{r_Z}.
\end{align}
For this PPOVM, the exclusion probabilities satisfy,
\begin{align}
    \text{Tr}[\Gamma_yT_x] &= \begin{cases}0,&y=x,\\ \frac{1}{2},&y\neq x,\end{cases} \\
    \sum_{y=1}^3\text{Tr}[\Gamma_yT_x] &= 1 \qquad \forall\ x.
\end{align}
Therefore, the ensemble admits strong-PAD. However, each pair contains a common Pauli unitary channel with weight $\frac{1}{2}$. For any probe, including one assisted by an auxiliary system, the two output states have a common nonzero positive contribution and cannot have orthogonal supports. Thus, the ensemble cannot be perfectly distinguished, even though it admits strong-PAD.
\end{example}

\begin{example}[\textbf{Pauli channels with full Pauli support}]
Consider a qubit Pauli channel  $\pcal^{\Vec{p}}$ with full Pauli support, that is, $p_{\alpha}>0$ for every $\alpha\in\{0,X,Y,Z\}$. Then, its Choi operator
    \(\Gamma^{\Vec{p}} = 2\sum_{\alpha}p_{\alpha}\dyad{r_{\alpha}} \)
satisfies 
\begin{align}
    \Gamma^{\Vec{p}} &\geq 2p_{\min}\,\id_A\otimes\id_B>0,
\end{align}
where $p_{\min}:=\min_{\alpha}p_{\alpha}>0$. Moreover, its four  eigenvalues are $2p_0,2p_X,2p_Y,2p_Z$, and thus it is positive definite. If this channel is a candidate of an ensemble, the PPOVM effect $T_x$ assigned to its exclusion must satisfy,
\begin{align}
    0=\text{Tr}[\Gamma^{\Vec{p}}T_x]\geq 2p_{\min}\text{Tr}[T_x].
\end{align}
Since $T_x\geq 0$, this forces $T_x=0$. Consequently, the outcome excluding this candidate never occurs for any member of the ensemble, and strong-PAD is impossible, in agreement with Proposition \ref{prop-oneChoiFullRank}. 

The presence of one such candidate does not, by itself, prevent weak-PAD. To illustrate this distinction, append the complete depolariser (which is a Pauli channel with full Pauli support) to the three-channel ensemble considered in the previous example. Its Pauli probabilities and Choi operator respectively are,
\begin{align}
    \Vec{p}^{(4)} &= \left(\frac{1}{4},\frac{1}{4},\frac{1}{4},\frac{1}{4}\right), ~~~~
    \Gamma_4 = \frac{1}{2}\id_A\otimes\id_B.
\end{align}
Now, retain the three PPOVM effects, $T_1,T_2,T_3$, given there and set $T_4=0$. Their normalisation is unchanged, and $\text{Tr}[\Gamma_xT_x]=0$ for all four candidates. Hence, the enlarged ensemble admits weak-PAD but cannot admit strong-PAD. Operationally, every observed outcome still excludes one of the first three candidates with certainty, while the completely depolarising channel is never excluded.

In contrast, if every candidate is a Pauli channel with full support, the same argument forces every PPOVM effect to vanish. This contradicts the normalisation condition and rules out even weak-PAD. A relevant instance is obtained by mixing each member of any Pauli-channel ensemble with a nonzero contribution from the complete depolariser. The resulting probabilities are,
\begin{align}
    \tilde{p}_{\alpha}^{(x)}=(1-\epsilon_x)p_{\alpha}^{(x)}+\frac{\epsilon_x}{4}>0,
\end{align}
for $0<\epsilon_x\leq 1$ and every $\alpha$. Therefore, even arbitrarily small positive admixtures of complete depolarisation to every candidate prevent weak-PAD. This also gives a  instance of the ``common-channel mixing'' obstruction in Proposition \ref{prop:mix-a-fixed-Choi}.
\end{example}

\begin{example}[\textbf{Replacer channels}]\label{example:commuting-replacer-channels}
Consider a Choi-commuting ensemble of replacer channels with replacement states $\{\nu_x\}_{x=1}^n$. Since $[\Gamma_x,\Gamma_y]=\id_A\otimes[\nu_x,\nu_y]$, this ensemble is Choi-commuting if and only if the replacement states themselves pairwise commute. Now, every PPOVM normalisation $\xi_A\otimes\id_B$  commutes with each candidate Choi operator, and consequently,
\begin{align}
    P_{\xcal}(\xi_A\otimes\id_B)=\xi_A\otimes\left(\prod_{x=1}^n\nu_x\right).
\end{align}
As $\xi_A$ is a state and hence nonzero, Theorem \ref{theorem:Choi-commuting-weak-PAD} translates to the following necessary and sufficient condition for weak-PAD, independent of the choice of PPOVM normalisation:
\begin{align}
    \prod_{x=1}^n\nu_x=0.
\end{align}
In other words, the commuting replacement states must have trivial common support. In their common eigenbasis, every basis vector must therefore have zero eigenvalue for at least one candidate state.

The independence from the PPOVM normalisation can also be seen as a consequence of Proposition \ref{prop:replacer-state<->channel}: the reduction of channel anti-distinguishability to state anti-distinguishability. For any PPOVM $T$, define the output POVM effects $M_x:=\text{Tr}_A[T_x]$. They satisfy,
\begin{align}
    \sum_x M_x &= \id_B, \\
    \text{Tr}[\Gamma_yT_x] &= \text{Tr}[\nu_yM_x].
\end{align}
Conversely, any POVM $M$ on the replacement states gives the same probabilities through $T_x=\xi_A\otimes M_x$, for any normalisation state $\xi_A$. Thus, both weak- and strong-PAD reduce to the corresponding conditions on the replacement states, and an unassisted probe suffices whenever perfect exclusion is possible. This reduction holds even without commutativity; the product criterion above uses the commutativity assumption.

For an explicit illustration, let the output system be a qutrit, with computational basis $\{\ket{0},\ket{1},\ket{2}\}$. Choose the replacement states to be
\begin{align}
    \nu_1 &= \frac{1}{2}\big(\dyad{1}+\dyad{2}\big), \\
    \nu_2 &= \frac{1}{2}\big(\dyad{0}+\dyad{2}\big), \\
    \nu_3 &= \frac{1}{2}\big(\dyad{0}+\dyad{1}\big).
\end{align}
These states commute and satisfy $\nu_1\nu_2\nu_3=0$. Measuring the output POVM with effects $M_1=\dyad{0}$, $M_2=\dyad{1}$ and $M_3=\dyad{2}$ gives,
\begin{align}
    \text{Tr}[\nu_yM_x] &= \begin{cases}0,&y=x,\\ \frac{1}{2},&y\neq x,\end{cases} \\
    \sum_{y=1}^3\text{Tr}[\nu_yM_x] &= 1 \qquad \forall\ x.
\end{align}
Consequently, the associated replacer channel ensemble admits strong-PAD. However, every pair of replacement states has overlapping supports and cannot be perfectly distinguished. An auxiliary system cannot remove this overlap: for any joint probe, the output is the product of its unchanged auxiliary marginal and the replacement state. Thus, these channels provide another instance of strong-PAD without perfect distinguishability.
\end{example}

\begin{example}[\textbf{Measure-and-prepare channels}]\label{example:commuting-measure-and-prepare}
 Consider an ensemble whose candidates share the von Neumann measurement $M_k=\dyad{k}_A$, but have different preparation sets $\{\varsigma_k^{(x)}\}_{k=0}^{d_A-1}$. Associated Choi operators are,
\begin{align}
    \Gamma_x &= \sum_{k=0}^{d_A-1}\dyad{k}_A\otimes\varsigma_k^{(x)}.
\end{align}
Suppose that, for each fixed measurement outcome $k$, the preparation states pairwise commute across the candidates, that is $[\varsigma_k^{(x)},\varsigma_k^{(y)}]=0$ for all $x,y$. The orthogonality of the input projectors then ensures that the Choi operators pairwise commute. For a diagonal normalisation state $\xi_A=\sum_k q_k\dyad{k}_A$, with $q_k\geq 0$ and $\sum_kq_k=1$, we have,
\begin{align}
    P_{\xcal}(\xi_A\otimes\id_B)=\sum_k q_k\dyad{k}_A\otimes\left(\prod_{x=1}^n\varsigma_k^{(x)}\right).
\end{align}
Since this normalisation also commutes with every candidate Choi operator, Theorem \ref{theorem:Choi-commuting-weak-PAD} gives that weak-PAD is possible with this normalisation if and only if,
\begin{align}
    \prod_{x=1}^n\varsigma_k^{(x)}=0 \qquad \text{for every }k\text{ with }q_k>0.
\end{align}
Thus, for every input basis state supported by the normalisation, the associated preparation states must satisfy the commuting replacer channel criterion of Example \ref{example:commuting-replacer-channels}. In particular, an unassisted probe $\dyad{k}_A$ suffices whenever this condition holds for that $k$.

To illustrate the dependence on the normalisation, consider channels with  qubit input and qutrit output. Let $\nu_1,\nu_2,\nu_3$ be the three commuting states considered in Example \ref{example:commuting-replacer-channels}. Define measure-and-prepare channel Choi operators
\begin{align}
    \Gamma_x &= \dyad{0}_A\otimes\nu_x+\dyad{1}_A\otimes\frac{\id_B}{3},~~\text{for } x=1,2,3.
\end{align}
These channels perform the same binary von Neumann measurement; outcome $0$ prepares the candidate-dependent state $\nu_x$, while outcome $1$ prepares the same maximally mixed state for all candidates. The Choi operators commute, and $\nu_1\nu_2\nu_3=0$ gives,
\begin{align}
    P_{\xcal}=\Gamma_1\Gamma_2\Gamma_3=\frac{1}{27}\dyad{1}_A\otimes\id_B.
\end{align}
Consequently, for an arbitrary normalisation state $\xi_A$, Proposition \ref{prop-CommutingCHoiNecessary} requires,
\begin{align}
    P_{\xcal}(\xi_A\otimes\id_B)=0
    \quad\Longleftrightarrow\quad \dyad{1}_A\xi_A=0.
\end{align}
Positivity and unit trace of $\xi_A$ imply that this is possible only for $\xi_A=\dyad{0}_A$. Hence, no positive definite normalisation state, including that of a maximally-assisted PPOVM, can admit even weak-PAD for this ensemble.

With the unassisted probe $\dyad{0}_A$, the three outputs are precisely $\nu_1,\nu_2,\nu_3$. The output measurement from Example \ref{example:commuting-replacer-channels} therefore gives the PPOVM effects,
\begin{align}
    T_x=\dyad{0}_A\otimes\dyad{x-1}_B,\qquad x=1,2,3.
\end{align}
They satisfy $\sum_xT_x=\dyad{0}_A\otimes\id_B$ and,
\begin{align}
    \text{Tr}[\Gamma_yT_x]=\begin{cases}0,&y=x,\\ \frac{1}{2},&y\neq x.\end{cases}
\end{align}
Every exclusion outcome has nonzero probability for the other two candidates, so the ensemble admits strong-PAD. This gives a Choi-commuting measure-and-prepare ensemble for which the choice of input state determines whether perfect exclusion is possible.
\end{example}
\section{Phase gates}\label{sec:phase-gates}
The exact criteria of Section \ref{sec:choi-commuting} rely on the pairwise commutativity of the candidate Choi operators. A different form of commutativity arises for unitary channels whose unitaries, rather than their Choi operators, pairwise commute. Since commuting unitaries are simultaneously diagonalisable, each such qubit unitary is, up to a global phase, a phase gate with respect to a common orthonormal eigenbasis. The Choi operators of these channels are rank-one, so that Proposition \ref{prop-oneChoiFullRank} does not obstruct perfect anti-distinguishability, and, as shown below, they do not commute in general, so that Theorem \ref{theorem:Choi-commuting-weak-PAD} does not apply either. Nevertheless, for this family, both weak- and strong-PAD admit exact characterisations that hold for arbitrary PPOVMs. 
\begin{definition}[\textbf{Qubit phase gate}]
Given an orthonormal basis $\{\ket{e_0},\ket{e_1} \}$ for a qubit, a $k$-th phase gate candidate is any diagonal unitary operator of the form:
\begin{align}
    U_k := e^{i\alpha_k}\dyad{e_0} + e^{i\beta_k}\dyad{e_1};~ \alpha_k, \beta_k \in [0,2\pi), \nonumber
\end{align}
with the associated unitary channel prescribed through the conjugation $\mathcal{U}^{(k)}(\cdot) = U_k (\cdot) U_k^{\dagger}$     
\end{definition}
Consider an ensemble $\xcal = \{\mathcal{U}^{(k)}\}_{k=1}^n$, with $n \ge 2$, whose associated unitaries share a common orthonormal eigenbasis $\{\ket{e_0}, \ket{e_1}\}$. Since the basis defining the Choi operator in Section \ref{sec:preliminaries} can be chosen without loss of generality, we take it to be the common eigenbasis, so that $\ket{\Gamma}_{AB} = \sum_{j} \ket{e_j}_A \otimes \ket{e_j}_B$. Defining the orthonormal vectors $\ket{f_j} := \ket{e_j}_A \otimes \ket{e_j}_B$ for $j = 0,1$, the Choi operators are, up to irrelevant global phases,
\begin{align}
    \Gamma_k = \dyad{g_k}, \qquad \ket{g_k} := \ket{f_0} + e^{i\Delta_k}\ket{f_1} \label{eq:phase-choi-vector},
\end{align}
where we define relative phases $\Delta_k := \beta_k - \alpha_k$. All candidate Choi operators are therefore supported on the two-dimensional subspace $\mathcal{K} := \text{span}\{\ket{f_0}, \ket{f_1}\}$, and, since $\braket{g_j}{g_k} = 1 + e^{i(\Delta_k - \Delta_j)}$, two of them commute only if their relative phases coincide or differ by $\pi$. The following proposition characterises perfect anti-distinguishability of such ensembles in terms of the relative phases $\Delta_k$ alone.

\begin{proposition}[\textbf{Exact conditions for qubit phase gate channels}]\label{prop:phase-gates}
    Let $\xcal$ be an ensemble of qubit phase-gate channels. 
    \begin{enumerate}
        \item $\xcal$ admits weak-PAD if and only if there exist weights $w_k \ge 0$, with $\sum_{k=1}^n w_k = 1$, such that
        \begin{align}
            \sum_{k=1}^{n} w_k e^{i\Delta_k} = 0. \label{eq:phase-weights}
        \end{align}
        \item $\xcal$ admits strong-PAD if and only if Eq.(\ref{eq:phase-weights}) holds with $w_k > 0$ for all $k$.
        \item Whenever weak- or strong-PAD is possible, it is achieved by an unassisted PPOVM with the input state $\ket{\psi} = (\ket{e_0} + \ket{e_1})/\sqrt{2}$.
    \end{enumerate}
\end{proposition}
\begin{proof}
    Let $P_{\mathcal{K}}$ denote the orthogonal projector onto $\mathcal{K}$.

    \textit{Necessity for weak-PAD.} Suppose that a PPOVM $T$, with normalisation $\xi_A \otimes \id_B$, weakly-PAD $\xcal$, and define the projected operators $C_k := P_{\mathcal{K}} T_k P_{\mathcal{K}} \ge 0$, that is, the restrictions of the PPOVM effects to $\mathcal{K}$. Since $\ket{g_k} \in \mathcal{K}$, the weak-PAD conditions read $\text{Tr}[\Gamma_k T_k] = \bra{g_k}C_k\ket{g_k} = 0$, and positivity gives $C_k\ket{g_k} = 0$. As $\mathcal{K}$ is two-dimensional, $C_k$ is supported on the orthogonal complement of $\ket{g_k}$ in $\mathcal{K}$, that is,
    \begin{align}
        C_k &= w_k \dyad{h_k}, \quad w_k \ge 0, \\ \ket{h_k} &:= \frac{1}{\sqrt{2}}\left(\ket{f_0} - e^{i\Delta_k}\ket{f_1}\right). \label{eq:phase-projected-effect}
    \end{align}
    Projecting both sides of the normalisation $\sum_k T_k = \xi_A \otimes \id_B$ onto $\mathcal{K}$ yields
    \begin{align}
        \sum_{k=1}^{n} C_k = P_{\mathcal{K}}(\xi_A \otimes \id_B)P_{\mathcal{K}}. \label{eq:phase-projected-normalisation}
    \end{align}
    Since $\bra{f_0}(\xi_A \otimes \id_B)\ket{f_1} = \bra{e_0}\xi_A\ket{e_1}\braket{e_0}{e_1} = 0$, the right-hand side is diagonal in the basis $\{\ket{f_0}, \ket{f_1}\}$, with diagonal entries $d_j := \bra{e_j}\xi_A\ket{e_j}$ satisfying $d_0 + d_1 = \text{Tr}[\xi_A] = 1$. In the same basis,
    \begin{align}
        C_k = \frac{w_k}{2}\begin{pmatrix} 1 & -e^{-i\Delta_k} \\ -e^{i\Delta_k} & 1 \end{pmatrix}.
    \end{align}
    Comparing the diagonal entries of Eq.(\ref{eq:phase-projected-normalisation}) gives $d_0 = d_1 = \frac{1}{2}\sum_k w_k$, and therefore $\sum_k w_k = 1$. Comparing the off-diagonal entries gives $\sum_k w_k e^{i\Delta_k} = 0$. Hence, Eq.(\ref{eq:phase-weights}) is necessary for weak-PAD, irrespective of the PPOVM.

    \textit{Sufficiency for weak-PAD.} Conversely, let $w_k \ge 0$, with $\sum_k w_k = 1$, satisfy Eq.(\ref{eq:phase-weights}). Use the unassisted input state $\ket{\psi} = (\ket{e_0} + \ket{e_1})/\sqrt{2}$, for which the PPOVM normalisation state is $\xi_A = \dyad{\psi}^T = \dyad{\psi}$, since $\ket{\psi}$ has real coefficients in this basis. Up to a global phase, the output of the $k$-th channel is $\ket{\phi_k} = (\ket{e_0} + e^{i\Delta_k}\ket{e_1})/\sqrt{2}$. With $\ket{\phi_k^{\perp}} := (\ket{e_0} - e^{i\Delta_k}\ket{e_1})/\sqrt{2}$, consider the output POVM effects
    \begin{align}
        M_k = 2w_k \dyad{\phi_k^{\perp}}. \label{eq:phase-output-povm}
    \end{align}
    In the basis $\{\ket{e_0}, \ket{e_1}\}$, $\sum_k M_k$ has diagonal entries $\sum_k w_k = 1$ and off-diagonal entries $-\sum_k w_k e^{\pm i\Delta_k} = 0$, so that $\sum_k M_k = \id_B$. The corresponding PPOVM effects
    \begin{align}
        T_k = \xi_A \otimes M_k \label{eq:phase-unassisted-ppovm}
    \end{align}
    are positive, satisfy $\sum_k T_k = \xi_A \otimes \id_B$, and give $\text{Tr}[\Gamma_k T_k] = \bra{\phi_k}M_k\ket{\phi_k} = 0$. Hence, $\xcal$ admits weak-PAD with an unassisted PPOVM.

    \textit{Strong-PAD.} By the above, every PPOVM that weakly-PAD $\xcal$ has projected effects $C_k$ of the form of Eq.(\ref{eq:phase-projected-effect}), with weights satisfying Eq.(\ref{eq:phase-weights}). For all $j$ and $k$,
    \begin{align}
        \text{Tr}[\Gamma_j T_k] &= \bra{g_j}C_k\ket{g_j} \\ &= w_k \abs{\braket{h_k}{g_j}}^2 \\ &= w_k\left[1 - \cos(\Delta_j - \Delta_k)\right],
    \end{align}
    and therefore
    \begin{align}
        \sum_{j=1}^{n} \text{Tr}[\Gamma_j T_k] = w_k \sum_{j=1}^{n}\left[1 - \cos(\Delta_j - \Delta_k)\right]. \label{eq:phase-strong-probability}
    \end{align}
    The same expression holds for the unassisted PPOVM of Eq.(\ref{eq:phase-unassisted-ppovm}), since $\bra{\phi_j}M_k\ket{\phi_j} = 2w_k\abs{\braket{\phi_k^{\perp}}{\phi_j}}^2 = w_k[1 - \cos(\Delta_j - \Delta_k)]$. Whenever Eq.(\ref{eq:phase-weights}) holds, the phase factors $e^{i\Delta_j}$ cannot all coincide, since a convex combination of identical unit-modulus numbers cannot vanish. Hence, for every $k$, there exists some $j$ with $e^{i\Delta_j} \neq e^{i\Delta_k}$, and the sum on the right-hand side of Eq.(\ref{eq:phase-strong-probability}) is strictly positive. Consequently, $\sum_j \text{Tr}[\Gamma_j T_k] > 0$ if and only if $w_k > 0$. Thus, strong-PAD requires Eq.(\ref{eq:phase-weights}) to hold with strictly positive weights and, conversely, if such weights exist, the unassisted PPOVM of Eq.(\ref{eq:phase-unassisted-ppovm}) establishes strong-PAD. This also proves statement 3.
\end{proof}

Proposition \ref{prop:phase-gates} gives a geometric characterisation of perfect anti-distinguishability for commuting qubit unitary channels. Each relative phase $\Delta_k$ determines a point $e^{i\Delta_k}$ on the unit circle. Weak-PAD is possible precisely when the origin lies in the convex hull of these points, whereas strong-PAD requires the origin to be a convex combination of all of them with strictly positive weights, that is, to lie in the relative interior of their convex hull. The proof also shows how the PPOVM normalisation enters: any weakly anti-distinguishing PPOVM must have a normalisation state that assigns equal weights $d_0 = d_1 = \frac{1}{2}$ to the eigenvectors $\ket{e_0}$ and $\ket{e_1}$. Both the unassisted input state $\ket{\psi}$ and the maximally-assisted normalisation $\frac{\id_A}{2} \otimes \id_B$ satisfy this requirement, and entanglement assistance offers no advantage for this family. For a fixed probe, the output states are pure states on a two-dimensional subspace, and the strong-PAD condition is consistent with the anti-distinguishability criterion for pure qubit states of Ref.~\cite{heinosaari2018antidistinguishability}, whose notion of anti-distinguishability requires every outcome to occur; the above proof additionally optimises over all PPOVMs and also treats weak-PAD. Finally, for $n = 2$, both conditions reduce to $e^{i\Delta_1} = -e^{i\Delta_2}$, which is the condition for perfect distinguishability of the pair, in agreement with Proposition \ref{prop-dist<=>AD_for2hypo}.

\begin{example}[\textbf{Trine phase-gate channels}]\label{example:trine-phase-gates}
    With $\omega = \exp(2\pi i/3)$ and $\{\ket{e_0}, \ket{e_1}\}$ the computational basis, consider the three phase gates $U_k = \dyad{0} + \omega^{k-1}\dyad{1}$, $k = 1,2,3$, with relative phases $\Delta_k = 2\pi(k-1)/3$. The uniform weights $w_1 = w_2 = w_3 = \frac{1}{3}$ satisfy
    \begin{align}
        \sum_{k=1}^{3} w_k e^{i\Delta_k} = \frac{1}{3}\left(1 + \omega + \omega^2\right) = 0.
    \end{align}
    Since all weights are strictly positive, Proposition \ref{prop:phase-gates} implies that the ensemble admits strong-PAD. Explicitly, with the unassisted input state $\ket{+} = (\ket{0} + \ket{1})/\sqrt{2}$, the output of the $k$-th channel is the trine state $\ket{\phi_k} = (\ket{0} + \omega^{k-1}\ket{1})/\sqrt{2}$, and the output POVM effects
    \begin{align}
        M_k = \frac{2}{3}\dyad{\phi_k^{\perp}}, \qquad \ket{\phi_k^{\perp}} = \frac{\ket{0} - \omega^{k-1}\ket{1}}{\sqrt{2}},
    \end{align}
    satisfy $\sum_k M_k = \id_B$ and $\bra{\phi_k}M_k\ket{\phi_k} = 0$. The corresponding PPOVM effects $T_k = \dyad{+}_A \otimes M_k$ satisfy $\sum_k T_k = \dyad{+}_A \otimes \id_B$ and $\text{Tr}[\Gamma_k T_k] = 0$.

    Moreover, by Eq.(\ref{eq:phase-strong-probability}),
    \begin{align}
        \text{Tr}[\Gamma_j T_k] = \begin{cases} 0, & j = k, \\ \frac{1}{2}, & j \neq k, \end{cases} \qquad \sum_{j=1}^{3}\text{Tr}[\Gamma_j T_k] = 1 \quad \forall\, k.
    \end{align}
    Thus, every exclusion outcome excludes its candidate with certainty and occurs with probability $\frac{1}{2}$ for each of the other two.
\end{example}
\section{Discussion}\label{sec:discussions}
Our results show that conclusively excluding a candidate quantum channel does not, in general, require perfect identification of the realised channel. For two channels, weak-PAD, strong-PAD and perfect distinguishability coincide. For larger ensembles, the implications from perfect distinguishability to strong-PAD and from strong-PAD to weak-PAD are strict. The Pauli and replacer-channel examples in Section \ref{sec:choi-commuting} demonstrate strong-PAD even when perfect distinguishability is impossible with auxiliary assistance. The distinction between weak- and strong-PAD is also operational: every observed outcome may exclude a candidate with certainty even though some candidates can never be excluded.

This latter distinction is reflected in the rank restrictions on the candidate Choi operators. Strong-PAD requires every candidate Choi operator to be rank-deficient, whereas weak-PAD requires at least two rank-deficient candidates. A positive definite Choi operator forces its exclusion effect to vanish, but does not by itself prevent weak-PAD of the ensemble. For example, appending the complete depolariser to the three-channel Pauli ensemble preserves weak-PAD while precluding strong-PAD. These restrictions also account for the obstructions obtained for measure-and-prepare channels with positive definite Choi operators and state-space contractions with positive definite noise states. Conversely, our union construction preserves weak- or strong-PAD when the constituent ensembles admit the corresponding form of exclusion. In particular, an ensemble can be embedded into a larger strong-PAD ensemble whenever a perfectly distinguishable partner channel exists for each of its candidates. Thus, the possibility of perfect exclusion depends on the ensemble to which a channel belongs, as well as on the channel itself.

The common Choi support imposes a further restriction that involves the testing strategy. By Theorem \ref{theorem-PAD_necessary}, this support must lie in the kernel of the PPOVM normalisation; a positive definite normalisation state therefore requires the common Choi support to be trivial. A rank-deficient normalisation can avoid this obstruction, as the measure-and-prepare examples demonstrate. In particular, Example \ref{example:commuting-measure-and-prepare} admits strong-PAD with an unassisted probe, whereas every positive definite normalisation state, including that of a maximally-assisted PPOVM, fails even for weak-PAD. Hence, the choice of probe remains relevant even when the candidate Choi operators commute.

Choi commutativity nevertheless allows the common-support restriction to become an exact criterion under an additional condition on the normalisation. When the PPOVM normalisation commutes with every candidate Choi operator, weak-PAD is possible precisely when the product of the Choi operators ``annihilates'' that normalisation. For qubit Pauli channels, this yields a criterion valid without restricting the testing strategy: each Pauli label must have zero probability for at least one candidate, and a maximally-assisted PPOVM suffices whenever this condition holds. For commuting replacer channels, the criterion reduces to a vanishing product of the replacement states and is independent of the normalisation. More generally, even without commutativity, both weak- and strong-PAD of replacer channels reduce exactly to the corresponding properties of their replacement states. For the measure-and-prepare family in Example \ref{example:commuting-measure-and-prepare}, the commuting criterion instead applies separately to the preparation states associated with each input basis state supported by a diagonal normalisation. These applications show both how the product condition simplifies the exclusion problem and why its consequences for probe choice differ between channel families.

A separate exact characterisation arises for the full permutation-channel ensemble associated with an unlabeled POVM: strong-PAD is possible if and only if at least one POVM effect is rank-deficient. The construction gives a direct operational interpretation. An unassisted probe chosen in the kernel of that effect makes its occurrence impossible, so an observed output label excludes every permutation assigning that effect to the observed label. The rank-one, von Neumann, trine and SIC examples in Section \ref{sec:measure-and-prepare} illustrate this procedure. Thus, the condition can be checked from the reference POVM itself, without testing each possible label assignment separately.

Some obstructions, however, cannot be avoided by choosing a different probe. Proposition \ref{prop:mix-a-fixed-Choi} rules out even weak-PAD whenever every candidate contains a nonzero admixed contribution from the same channel. This explains both the impossibility of weak-PAD for ensembles of state-space contractions retaining a nonzero identity contribution in every candidate and the loss of weak-PAD when every Pauli candidate is mixed with a positive contribution from complete depolarisation. The same result applies when an incompatible ensemble is made compatible by adding a nonzero contribution from the same noise channel to every candidate, as in Example \ref{example:incompatibility-broken}. The obstruction already holds for any positive mixing weights, whether or not they suffice to achieve compatibility. It therefore does not imply that compatibility alone prevents perfect anti-distinguishability.

Beyond the Choi-commuting ensembles, Section \ref{sec:phase-gates} provides an exact characterisation of perfect anti-distinguishability for commuting qubit unitary channels. Although the corresponding unitaries share a common eigenbasis, their Choi operators do not generally commute, so that the criterion of Section \ref{sec:choi-commuting} does not directly apply. Instead, weak-PAD is possible precisely when the origin lies in the convex hull of the unit-circle points determined by the relative phases, whereas strong-PAD requires the origin to lie in the relative interior of their convex hull. Moreover, we find that whenever these conditions are satisfied, they can be realised by an unassisted probe. Thus, entanglement assistance offers no advantage for the feasibility of perfect anti-distinguishability within this family. This focused characterisation complements the dedicated study of unitary channel anti-distinguishability by Manna and Das Bhowmik \cite{manna2026unitaries}, without undertaking a systematic and complete study of arbitrary unitary ensembles.

\textbf{Multiple queries and memory channels.} The problem statements formulated, for single-queried quantum channels, using PPOVM can be translated without much effort to multiple-queried quantum channels and even memory channels \cite{Kretschmann2005QuantumMemory} (multiple-time-step quantum processes where the steps are causally correlated in time; thus even the previous multiple-queried scenario can be considered as one without any correlation in time). Choi operators of such processes are described by quantum combs \cite{Chiribella2008_PRL,Chiribella2008_EPL_supermaps,Chiribella2009_PRA_networks} and measurement strategies of these combs are described by quantum testers \cite{Chiribella2008_memory_discrimination} which are higher-order generalisations of PPOVMs. Structurally, most of our no-go results should easily extend to these higher-order settings. Especially, obstructions due to strict positive definiteness of the Choi operator should signal identical obstructions if the involved quantum combs are  positive definite.  

\textbf{Future directions.} 
The scope of our work was restricted to perfect anti-distinguishability and primarily characterisation of channels and ensembles which allow or disallow this. This enabled us to talk about how candidature of a single specific channel could obstruct perfect anti-distinguishability or how addition of a  channel could activate already disallowed anti-distinguishability. Future studies could extend to minimum-error performance analyses of given ensembles. This might also be a regime where multiple-queried quantum channels prove to improve anti-distinguishability performances. Considering more general quantum combs, how associated anti-distinguishability is affected by inherent non-Markovianity or temporal correlations can be studied. Moreover, for a fixed ensemble of quantum combs, one can study the performance gaps established by different classes of measurement strategies, for example, how much does adaptivity or quality of quantum memories within a strategy contribute to the overall performance. Another avenue is to determine whether certain ensembles, not perfectly distinguishable, can be unambiguously anti-distinguished by admitting an inconclusive measurement outcome. Furthermore, since we have a weaker and stronger notion of anti-distinguishability, one could very well attempt to characterise the implications of these notions to unambiguous anti-distinguishability.

\begin{acknowledgements}
    A.A.V. acknowledges support from QUAS VEGA 2/0164/25. N.S.R. acknowledges financial support from the US Department of Energy, Office of Science, Advanced Scientific Computing Research program, under award number DE-SC0025430.
The authors used ChatGPT (OpenAI) and Claude (Anthropic) to assist with exploratory algebra, language editing, manuscript organisation, and consistency checks. The authors take full responsibility for the scientific content, proofs, references, and final presentation of this work.
\end{acknowledgements}

\onecolumngrid
\appendix*
\section{Perfectly distinguishable channel pairs} \label{appendix-PDchannelsPairs}
It is a well known fact that, given a rank deficient quantum state $\varrho$  one can always find another state $\psi$ such that $\varrho \psi =0$. This also translates to the fact that such a $\varrho$ can always have a partner state $\psi$ such that the  pair is perfectly distinguishable with a single copy \cite{heinosaari2011mathematical}. For quantum channels, however, we are not aware of an explicit construction of such a partner for a given channel with a rank-deficient Choi operator. In the light of this, we ask the question: 

\vspace{0.5cm}
``Given the Choi operator of a quantum channel, can one find at least one another channel, and a probe state (or equivalently, PPOVM normalisation),  such that the pair can be perfectly distinguished with the accompanying probe?''

\vspace{0.5cm}
Addressing this leads to  following proposition.
\begin{proposition}[\textbf{Perfect distinguishability partner}]\label{prop:PD-partner}
    Let $\ecal_{A \rightarrow B}$ be a quantum channel with
    Choi operator $\Gamma^{\ecal}_{AB}$. Then there exists
    a quantum channel $\fcal_{A \rightarrow B}$ such that
    the pair $\{\ecal_{A \rightarrow B},\fcal_{A \rightarrow B}\}$
    is perfectly distinguishable in the single-query assisted
    scenario if and only if $\Gamma^{\ecal}_{AB}$ is
    rank-deficient.
\end{proposition}

\begin{proof}
    Recall that two quantum channels with respective Choi
    operators $\Gamma_a$ and $\Gamma_b$ are perfectly
    distinguishable in the single-query assisted scenario
    if and only if there exists a quantum state $\rho_A$
    such that \cite{Chiribella2008_memory_discrimination}
    \begin{equation}
        \Gamma_a(\rho_A \otimes \id_B)\Gamma_b = 0.
        \label{eq:PD_criterion}
    \end{equation}
    Here, $\rho_A$ is the normalisation state of the
    corresponding PPOVM. We prove both directions.

    \medskip
    \noindent
    \textit{Sufficiency.}
    Suppose that $\Gamma^{\ecal}_{AB}$ is rank-deficient,
    i.e., $\ker(\Gamma^{\ecal}_{AB}) \neq \{0\}$.
    Our task is to construct a valid Choi operator
    $\Gamma^{\fcal}_{AB}$ and a normalisation state $\rho_A$
    satisfying Eq.~\eqref{eq:PD_criterion}.

    Let $\Pi_{AB}$ denote the orthogonal projector onto
    $\ker(\Gamma^{\ecal}_{AB})$:
    \begin{equation}
        \Pi_{AB}
        := \Pi_{\ker(\Gamma^{\ecal}_{AB})}.
    \end{equation}
    Since $\Pi_{AB}$ is supported entirely within the
    kernel of $\Gamma^{\ecal}_{AB}$, we have
    \begin{equation}
        \Gamma^{\ecal}_{AB}\Pi_{AB}=0.
        \label{eq:GammaPi}
    \end{equation}

    Define the positive operator
    \begin{equation}
        \Xi_A := \text{Tr}_B[\Pi_{AB}].
    \end{equation}
    Since $\Gamma^{\ecal}_{AB}$ is rank-deficient,
    $\Pi_{AB}\neq 0$, and therefore
    \begin{equation}
        \text{Tr}[\Xi_A]
        =\text{Tr}[\Pi_{AB}]
        =\dim\ker(\Gamma^{\ecal}_{AB})>0.
    \end{equation}
    Consequently, $\Xi_A$ is a nonzero positive operator.

    Let $S:=\text{supp}(\Xi_A)$, and denote by $P_S$
    and $P_{S^\perp}$ the orthogonal projectors onto
    $S$ and $S^\perp=\ker(\Xi_A)$, respectively.
    Define $\Xi_A^{-1/2}$ as the Moore--Penrose
    pseudoinverse square root of $\Xi_A$, namely,
    the inverse of $\Xi_A^{1/2}$ on $S$ and zero
    on $S^\perp$. Thus,
    \begin{equation}
        \Xi_A^{-1/2}\Xi_A\Xi_A^{-1/2}=P_S.
        \label{eq:Xi_pseudoinverse}
    \end{equation}

    We first establish the identity
    \begin{equation}
        (P_S\otimes\id_B)\Pi_{AB}=\Pi_{AB}.
        \label{eq:PSPi}
    \end{equation}
    Since $P_{S^\perp}$ is the orthogonal projector
    onto $\ker(\Xi_A)$ and $\Xi_A$ is Hermitian,
    we have
    \begin{equation}
        P_{S^\perp}\Xi_A
        =\Xi_A P_{S^\perp}=0.
    \end{equation}
    Consequently,
    \begin{equation}
        \text{Tr}[(P_{S^\perp}\otimes\id_B)\Pi_{AB}]
        =\text{Tr}[P_{S^\perp}\Xi_A]=0.
    \end{equation}
    Define $Q_{AB}:=P_{S^\perp}\otimes\id_B$.
    Since both $Q_{AB}$ and $\Pi_{AB}$ are
    orthogonal projectors, we have
    $Q_{AB}^2=Q_{AB}$ and $\Pi_{AB}^2=\Pi_{AB}$.
    Therefore,
    \begin{align}
        0
        &=\text{Tr}[Q_{AB}\Pi_{AB}] \nonumber\\
        &=\text{Tr}[\Pi_{AB}Q_{AB}\Pi_{AB}]
          \nonumber\\
        &=\text{Tr}[
            (Q_{AB}\Pi_{AB})^\dagger
            (Q_{AB}\Pi_{AB})] \nonumber\\
        &=\|Q_{AB}\Pi_{AB}\|_2^2,
    \end{align}
    where $\|\cdot\|_2$ denotes the Hilbert--Schmidt
    norm. Since this norm vanishes only for the
    zero operator, we obtain
    \begin{equation}
        (P_{S^\perp}\otimes\id_B)\Pi_{AB}=0.
    \end{equation}
    Hence,
    \begin{align}
        \Pi_{AB}
        &=\bigl((P_S+P_{S^\perp})\otimes\id_B\bigr)
          \Pi_{AB}\nonumber\\
        &=(P_S\otimes\id_B)\Pi_{AB},
    \end{align}
    establishing Eq.~\eqref{eq:PSPi}.

    Now, define the operator
    \begin{align}
        \Gamma^{\fcal}_{AB}
        :=\;&(\Xi_A^{-1/2}\otimes\id_B)\Pi_{AB}
             (\Xi_A^{-1/2}\otimes\id_B)
             \nonumber\\
        &+\frac{1}{d_B}P_{S^\perp}\otimes\id_B,
        \label{eq:GammaF_partner}
    \end{align}
    where $d_B:=\dim(\h_B)$.
    Both summands are positive semidefinite,
    and hence $\Gamma^{\fcal}_{AB}\geq 0$.

    We now verify the Choi normalisation condition.
    Taking the partial trace over $B$, we obtain
    \begin{align}
        \text{Tr}_B[\Gamma^{\fcal}_{AB}]
        &=\Xi_A^{-1/2}\text{Tr}_B[\Pi_{AB}]
          \Xi_A^{-1/2}
          +P_{S^\perp}\nonumber\\
        &=\Xi_A^{-1/2}\Xi_A\Xi_A^{-1/2}
          +P_{S^\perp}\nonumber\\
        &=P_S+P_{S^\perp}\nonumber\\
        &=\id_A,
    \end{align}
    where we used Eq.~\eqref{eq:Xi_pseudoinverse}.
    Therefore, $\Gamma^{\fcal}_{AB}$ is the
    Choi operator of a valid quantum channel
    $\fcal_{A\rightarrow B}$.

    Next, choose the PPOVM normalisation state
    \begin{equation}
        \rho_A
        :=\frac{\Xi_A^{1/2}}
                {\text{Tr}[\Xi_A^{1/2}]}.
        \label{eq:xi_partner}
    \end{equation}
    Since $\Xi_A$ is positive semidefinite and
    nonzero, $\text{Tr}[\Xi_A^{1/2}]>0$.
    Consequently, $\rho_A\geq 0$ and
    $\text{Tr}[\rho_A]=1$, confirming that
    $\rho_A$ is a valid density operator.
    Moreover, $\text{supp}(\rho_A)=S$, so that
    \begin{equation}
        \rho_A P_{S^\perp}=0.
    \end{equation}

    We now show that the constructed channel
    $\fcal_{A\rightarrow B}$ and normalisation
    state $\rho_A$ satisfy the perfect
    distinguishability condition.
    Using Eq.~\eqref{eq:GammaF_partner} and
    $\rho_A P_{S^\perp}=0$, we have
    \begin{align}
        &(\rho_A\otimes\id_B)\Gamma^{\fcal}_{AB}
        \nonumber\\
        &=(\rho_A\Xi_A^{-1/2}\otimes\id_B)
          \Pi_{AB}
          (\Xi_A^{-1/2}\otimes\id_B).
    \end{align}
    From Eq.~\eqref{eq:xi_partner},
    \begin{align}
        \rho_A\Xi_A^{-1/2}
        &=\frac{\Xi_A^{1/2}\Xi_A^{-1/2}}
                {\text{Tr}[\Xi_A^{1/2}]}
          \nonumber\\
        &=\frac{P_S}{\text{Tr}[\Xi_A^{1/2}]}.
    \end{align}
    Therefore,
    \begin{align}
        &(\rho_A\otimes\id_B)\Gamma^{\fcal}_{AB}
        \nonumber\\
        &=\frac{(P_S\otimes\id_B)\Pi_{AB}
                (\Xi_A^{-1/2}\otimes\id_B)}
               {\text{Tr}[\Xi_A^{1/2}]}
          \nonumber\\
        &=\frac{\Pi_{AB}
                (\Xi_A^{-1/2}\otimes\id_B)}
               {\text{Tr}[\Xi_A^{1/2}]},
    \end{align}
    where the last equality follows from
    Eq.~\eqref{eq:PSPi}.

    Multiplying by $\Gamma^{\ecal}_{AB}$
    from the left and using Eq.~\eqref{eq:GammaPi},
    we obtain
    \begin{align}
        &\Gamma^{\ecal}_{AB}
         (\rho_A\otimes\id_B)\Gamma^{\fcal}_{AB}
         \nonumber\\
        &=\frac{\Gamma^{\ecal}_{AB}\Pi_{AB}
                (\Xi_A^{-1/2}\otimes\id_B)}
               {\text{Tr}[\Xi_A^{1/2}]}
          \nonumber\\
        &=0.
    \end{align}
    Thus, the pair
    $\{\ecal_{A\rightarrow B},\fcal_{A\rightarrow B}\}$
    is perfectly distinguishable in the
    single-query assisted scenario.

    \medskip
    \noindent
    \textit{Necessity.}
    Conversely, suppose there exists a quantum channel
    $\fcal_{A\rightarrow B}$ such that the pair
    $\{\ecal_{A\rightarrow B},\fcal_{A\rightarrow B}\}$
    is perfectly distinguishable in the single-query
    assisted scenario.
    By Eq.~\eqref{eq:PD_criterion}, there exists a
    normalisation state $\rho_A$ satisfying
    \begin{equation}
        \Gamma^{\ecal}_{AB}
        (\rho_A\otimes\id_B)
        \Gamma^{\fcal}_{AB}=0.
        \label{eq:PD_necessity}
    \end{equation}

    Suppose, for contradiction, that
    $\Gamma^{\ecal}_{AB}$ is full-rank.
    Since $\Gamma^{\ecal}_{AB}\geq 0$,
    it is then positive definite and hence invertible.
    Multiplying Eq.~\eqref{eq:PD_necessity}
    from the left by $(\Gamma^{\ecal}_{AB})^{-1}$
    gives
    \begin{equation}
        (\rho_A\otimes\id_B)\Gamma^{\fcal}_{AB}=0.
    \end{equation}
    Taking the trace of both sides,
    \begin{align}
        0
        &=\text{Tr}[
            (\rho_A\otimes\id_B)\Gamma^{\fcal}_{AB}]
          \nonumber\\
        &=\text{Tr}[
            \rho_A\text{Tr}_B[\Gamma^{\fcal}_{AB}]]
          \nonumber\\
        &=\text{Tr}[\rho_A\id_A]\nonumber\\
        &=1,
    \end{align}
    where we used the Choi normalisation
    $\text{Tr}_B[\Gamma^{\fcal}_{AB}]=\id_A$
    and $\text{Tr}[\rho_A]=1$.

    This is a contradiction. Therefore,
    $\Gamma^{\ecal}_{AB}$ cannot be full-rank,
    implying
    \begin{equation}
        \ker(\Gamma^{\ecal}_{AB})\neq\{0\}.
    \end{equation}

    Combining both directions completes the proof.
\end{proof}

\bibliographystyle{quantum}
\bibliography{bibliography}

\end{document}

%% file: preamble.tex
\usepackage[utf8]{inputenc}

\usepackage{amsmath,amssymb,amsfonts,amsthm,mathtools,cases}
\usepackage{array}
\usepackage{mathrsfs}
\usepackage{bm} 
\usepackage{physics}
\usepackage{leftindex,tensor}

\usepackage{graphicx}
\usepackage{tcolorbox} 

\usepackage{color}
\usepackage{xcolor} 

\usepackage{hyperref} 

\hypersetup{
	colorlinks  = true, 
	linkcolor   = magenta, 
	citecolor   = magenta, 
	filecolor   = magenta, 
	urlcolor    = blue} 

\usepackage{dsfont}
\usepackage{bbm} 

\usepackage{ulem} 
\usepackage{epigraph,marginnote}
\usepackage{enumitem} 
\usepackage{appendix}
\usepackage{soul} 
\usepackage{orcidlink} 

\theoremstyle{definition}
\newtheorem{proposition}{Proposition}
\newtheorem{example}{Example}
\newtheorem{theorem}{Theorem}
\newtheorem{lemma}{Lemma}
\newtheorem{definition}{Definition}
\newtheorem{coro}{Corollary}
\newtheorem{remark}{Remark}

\newcommand{\ccal}{\mathcal{C}}  
\newcommand{\ecal}{\mathcal{E}}  
\newcommand{\fcal}{\mathcal{F}}  

\newcommand{\h}{\mathcal{H}}     

\newcommand{\mcal}{\mathcal{M}}  
\newcommand{\pcal}{\mathcal{P}}  
\newcommand{\scal}{\mathcal{S}}  
\newcommand{\xcal}{\mathcal{X}}  

\newcommand{\half}{\tfrac{1}{2}}

\newcommand{\id}{\mathbbm{1}} 



%% file: bibliography.bib
@article{Jamiolkowski1972,
  title         = {Linear transformations which preserve trace and positive semidefiniteness of operators},
  author        = {Jamio{\l}kowski, Andrzej},
  journal       = {Reports on Mathematical Physics},
  volume        = {3},
  number        = {4},
  pages         = {275--278},
  year          = {1972},
  doi           = {10.1016/0034-4877(72)90011-0},
  url           = {https://doi.org/10.1016/0034-4877(72)90011-0}
}

@article{Pusey2012Reality,
  title     = {On the reality of the quantum state},
  author    = {Pusey, Matthew F. and Barrett, Jonathan and Rudolph, Terry},
  journal   = {Nature Physics},
  volume    = {8},
  pages     = {475--478},
  year      = {2012},
  doi       = {10.1038/nphys2309}
}

@article{Choi1975,
  title         = {Completely positive linear maps on complex matrices},
  author        = {Choi, Man-Duen},
  journal       = {Linear Algebra and its Applications},
  volume        = {10},
  number        = {3},
  pages         = {285--290},
  year          = {1975},
  doi           = {10.1016/0024-3795(75)90075-0},
  url           = {https://doi.org/10.1016/0024-3795(75)90075-0}
}

@article{ziman2008process,
  title         = {Process positive-operator-valued measure: A mathematical framework for the description of process tomography experiments},
  author        = {Ziman, M{\'a}rio},
  journal       = {Physical Review A},
  volume        = {77},
  number        = {6},
  pages         = {062112},
  year          = {2008},
  doi           = {10.1103/PhysRevA.77.062112},
  url           = {https://doi.org/10.1103/PhysRevA.77.062112},
  eprint        = {0802.3862},
  archivePrefix = {arXiv},
  primaryClass  = {quant-ph}
}

@article{Chiribella2008_PRL,
  author        = {Chiribella, Giulio and D'Ariano, Giacomo Mauro and Perinotti, Paolo},
  title         = {Quantum circuit architecture},
  journal       = {Physical Review Letters},
  volume        = {101},
  number        = {6},
  pages         = {060401},
  year          = {2008},
  doi           = {10.1103/PhysRevLett.101.060401},
  url           = {https://doi.org/10.1103/PhysRevLett.101.060401},
  eprint        = {0712.1325},
  archivePrefix = {arXiv},
  primaryClass  = {quant-ph}
}

@article{Chiribella2008_memory_discrimination,
  author        = {Chiribella, Giulio and D'Ariano, Giacomo Mauro and Perinotti, Paolo},
  title         = {Memory effects in quantum channel discrimination},
  journal       = {Physical Review Letters},
  volume        = {101},
  pages         = {180501},
  year          = {2008},
  doi           = {10.1103/PhysRevLett.101.180501},
  url           = {https://doi.org/10.1103/PhysRevLett.101.180501},
  eprint        = {0803.3237},
  archivePrefix = {arXiv},
  primaryClass  = {quant-ph}
}

@article{Chiribella2008_EPL_supermaps,
  author        = {Chiribella, Giulio and D'Ariano, Giacomo Mauro and Perinotti, Paolo},
  title         = {Transforming quantum operations: Quantum supermaps},
  journal       = {Europhysics Letters},
  volume        = {83},
  pages         = {30004},
  year          = {2008},
  doi           = {10.1209/0295-5075/83/30004},
  url           = {https://doi.org/10.1209/0295-5075/83/30004},
  eprint        = {0804.0180},
  archivePrefix = {arXiv},
  primaryClass  = {quant-ph}
}

@article{Chiribella2009_PRA_networks,
  author        = {Chiribella, Giulio and D'Ariano, Giacomo Mauro and Perinotti, Paolo},
  title         = {Theoretical framework for quantum networks},
  journal       = {Physical Review A},
  volume        = {80},
  number        = {2},
  pages         = {022339},
  year          = {2009},
  doi           = {10.1103/PhysRevA.80.022339},
  url           = {https://doi.org/10.1103/PhysRevA.80.022339},
  eprint        = {0904.4483},
  archivePrefix = {arXiv},
  primaryClass  = {quant-ph}
}

@article{stratton2024operational,
  title         = {Operational interpretation of the Choi rank through exclusion tasks},
  author        = {Stratton, Benjamin and Hsieh, Chung-Yun and Skrzypczyk, Paul},
  journal       = {Physical Review A},
  volume        = {110},
  number        = {5},
  pages         = {L050601},
  year          = {2024},
  doi           = {10.1103/PhysRevA.110.L050601},
  url           = {https://doi.org/10.1103/PhysRevA.110.L050601},
  eprint        = {2406.08360},
  archivePrefix = {arXiv},
  primaryClass  = {quant-ph}
}

@article{ragini2024higher,
  title         = {Higher-order incompatibility improves distinguishability of causal quantum networks},
  author        = {Sudarsanan Ragini, Nidhin and Sazim, Sk},
  journal       = {New Journal of Physics},
  volume        = {26},
  number        = {12},
  pages         = {123003},
  year          = {2024},
  doi           = {10.1088/1367-2630/ad93f6},
  url           = {https://doi.org/10.1088/1367-2630/ad93f6},
  eprint        = {2405.20080},
  archivePrefix = {arXiv},
  primaryClass  = {quant-ph}
}

@article{uola2019quantifying,
  title         = {Quantifying quantum resources with conic programming},
  author        = {Uola, Roope and Kraft, Tristan and Shang, Jiangwei and Yu, Xiao-Dong and G{\"u}hne, Otfried},
  journal       = {Physical Review Letters},
  volume        = {122},
  number        = {13},
  pages         = {130404},
  year          = {2019},
  doi           = {10.1103/PhysRevLett.122.130404},
  url           = {https://doi.org/10.1103/PhysRevLett.122.130404},
  eprint        = {1812.09216},
  archivePrefix = {arXiv},
  primaryClass  = {quant-ph}
}

@book{heinosaari2011mathematical,
  title     = {The Mathematical Language of Quantum Theory: From Uncertainty to Entanglement},
  author    = {Heinosaari, Teiko and Ziman, M{\'a}rio},
  publisher = {Cambridge University Press},
  year      = {2011},
  doi       = {10.1017/CBO9781139031103},
  url       = {https://doi.org/10.1017/CBO9781139031103},
  isbn      = {9780521195836}
}

@article{bandyopadhyay2014conclusive,
  title         = {Conclusive exclusion of quantum states},
  author        = {Bandyopadhyay, Somshubhro and Jain, Rahul and Oppenheim, Jonathan and Perry, Christopher},
  journal       = {Physical Review A},
  volume        = {89},
  number        = {2},
  pages         = {022336},
  year          = {2014},
  doi           = {10.1103/PhysRevA.89.022336},
  url           = {https://doi.org/10.1103/PhysRevA.89.022336},
  eprint        = {1306.4683},
  archivePrefix = {arXiv},
  primaryClass  = {quant-ph}
}

@article{caves2002conditions,
  title         = {Conditions for compatibility of quantum-state assignments},
  author        = {Caves, Carlton M. and Fuchs, Christopher A. and Schack, R{\"u}diger},
  journal       = {Physical Review A},
  volume        = {66},
  number        = {6},
  pages         = {062111},
  year          = {2002},
  doi           = {10.1103/PhysRevA.66.062111},
  url           = {https://doi.org/10.1103/PhysRevA.66.062111},
  eprint        = {quant-ph/0206110},
  archivePrefix = {arXiv},
  primaryClass  = {quant-ph}
}

@article{heinosaari2024simple,
  title         = {Simple information-processing tasks with unbounded quantum advantage},
  author        = {Heinosaari, Teiko and Kerppo, Oskari and Lepp{\"a}j{\"a}rvi, Leevi and Pl{\'a}vala, Martin},
  journal       = {Physical Review A},
  volume        = {109},
  number        = {3},
  pages         = {032627},
  year          = {2024},
  doi           = {10.1103/PhysRevA.109.032627},
  url           = {https://doi.org/10.1103/PhysRevA.109.032627},
  eprint        = {2308.07727},
  archivePrefix = {arXiv},
  primaryClass  = {quant-ph}
}

@article{leifer2020noncontextuality,
  title         = {Noncontextuality inequalities from antidistinguishability},
  author        = {Leifer, Matthew and Duarte, Cristina},
  journal       = {Physical Review A},
  volume        = {101},
  number        = {6},
  pages         = {062113},
  year          = {2020},
  doi           = {10.1103/PhysRevA.101.062113},
  url           = {https://doi.org/10.1103/PhysRevA.101.062113},
  eprint        = {2001.11485},
  archivePrefix = {arXiv},
  primaryClass  = {quant-ph}
}

@article{russo2023inner,
  title         = {Inner products of pure states and their antidistinguishability},
  author        = {Russo, Vincent and Sikora, Jamie},
  journal       = {Physical Review A},
  volume        = {107},
  pages         = {L030202},
  year          = {2023},
  doi           = {10.1103/PhysRevA.107.L030202},
  url           = {https://doi.org/10.1103/PhysRevA.107.L030202},
  eprint        = {2206.08313},
  archivePrefix = {arXiv},
  primaryClass  = {quant-ph}
}

@article{srikumar2026contextuality,
  title         = {How Contextuality and
                   Antidistinguishability Are Related},
  author        = {Srikumar, Maiyuren and Bartlett, Stephen D.
                   and Karanjai, Angela},
  journal       = {Physical Review Letters},
  volume        = {136},
  number        = {8},
  pages         = {080203},
  year          = {2026},
  doi           = {10.1103/tlh1-xwks},
  eprint        = {2411.09919},
  archivePrefix = {arXiv},
  primaryClass  = {quant-ph},
  url           = {https://doi.org/10.1103/tlh1-xwks}
}

@article{mcirvin2024quantum,
  title         = {Quantum state exclusion through offset measurement},
  author        = {McIrvin, Caleb and Mohan, Ankith and Sikora, Jamie},
  journal       = {Physical Review A},
  volume        = {110},
  number        = {4},
  pages         = {042211},
  year          = {2024},
  doi           = {10.1103/PhysRevA.110.042211},
  url           = {https://doi.org/10.1103/PhysRevA.110.042211},
  eprint        = {2403.17252},
  archivePrefix = {arXiv},
  primaryClass  = {quant-ph}
}

@article{johnston2025tight,
  title         = {Tight bounds for antidistinguishability and circulant sets of pure quantum states},
  author        = {Johnston, Nathaniel and Russo, Vincent and Sikora, Jamie},
  journal       = {Quantum},
  volume        = {9},
  pages         = {1622},
  year          = {2025},
  doi           = {10.22331/q-2025-02-04-1622},
  url           = {https://doi.org/10.22331/q-2025-02-04-1622},
  eprint        = {2311.17047},
  archivePrefix = {arXiv},
  primaryClass  = {quant-ph}
}

@article{hillery2025group,
  title   = {A group-theoretic approach to elimination measurements of qubit sequences},
  author  = {Hillery, Mark and Andersson, Erika and Puthoor, Ittoop V.},
  journal = {Journal of Physics A: Mathematical and Theoretical},
  volume  = {58},
  number  = {42},
  pages   = {425301},
  year    = {2025},
  doi     = {10.1088/1751-8121/ae1079},
  url     = {https://doi.org/10.1088/1751-8121/ae1079}
}

@article{mannaPRAmeasurments,
  title         = {Single-shot distinguishability and antidistinguishability of quantum measurements},
  author        = {Manna, Satyaki and Suresh, Sneha and Kachhawaha, Manan Singh and Saha, Debashis},
  journal       = {Physical Review A},
  volume        = {111},
  number        = {2},
  pages         = {022221},
  year          = {2025},
  doi           = {10.1103/PhysRevA.111.022221},
  url           = {https://doi.org/10.1103/PhysRevA.111.022221},
  eprint        = {2410.10632},
  archivePrefix = {arXiv},
  primaryClass  = {quant-ph}
}

@article{manna2026unitaries,
  title         = {Single-shot antidistinguishability
                   of unitary operations},
  author        = {Manna, Satyaki and Das Bhowmik, Anandamay},
  journal       = {Physical Review A},
  volume        = {113},
  number        = {2},
  pages         = {022218},
  year          = {2026},
  doi           = {10.1103/d183-k1x3},
  eprint        = {2510.14609},
  archivePrefix = {arXiv},
  primaryClass  = {quant-ph},
  url           = {https://doi.org/10.1103/d183-k1x3}
}

@misc{manna-entagldstatesForMeasurements,
  title         = {All $d\otimes d$ dimensional entangled states are useful for the antidiscrimination of quantum measurements when $d$ is even},
  author        = {Manna, Satyaki},
  year          = {2025},
  eprint        = {2510.26255},
  archivePrefix = {arXiv},
  primaryClass  = {quant-ph},
  doi           = {10.48550/arXiv.2510.26255},
  url           = {https://arxiv.org/abs/2510.26255}
}

@article{yao2026conclusive,
  title         = {Conclusive exclusion of quantum states
                   with group action},
  author        = {Yao, Hongshun and Wang, Xin},
  journal       = {Physical Review A},
  volume        = {113},
  number        = {2},
  pages         = {022205},
  year          = {2026},
  doi           = {10.1103/bxn4-4v6p},
  eprint        = {2503.04605},
  archivePrefix = {arXiv},
  primaryClass  = {quant-ph},
  url           = {https://doi.org/10.1103/bxn4-4v6p}
}

@article{diebra2026quantum,
  title         = {Quantum state exclusion for group-generated
                   ensembles of pure states},
  author        = {Diebra, Arnau and Llorens, Santiago
                   and Bagan, Emili and Sent{\'i}s, Gael
                   and Mu{\~n}oz-Tapia, Ramon},
  journal       = {Physical Review Research},
  volume        = {8},
  number        = {1},
  pages         = {L012001},
  year          = {2026},
  doi           = {10.1103/2k5d-bprn},
  eprint        = {2503.02568},
  archivePrefix = {arXiv},
  primaryClass  = {quant-ph},
  url           = {https://doi.org/10.1103/2k5d-bprn}
}

@article{heinosaari2018antidistinguishability,
  title={Antidistinguishability of pure quantum states},
  author={Heinosaari, Teiko and Kerppo, Oskari},
  journal={Journal of Physics A: Mathematical and Theoretical},
  volume={51},
  number={36},
  pages={365303},
  year={2018},
  publisher={IOP Publishing},
  doi={10.1088/1751-8121/aad1fc},
  url={https://doi.org/10.1088/1751-8121/aad1fc},
  eprint={1804.10457},
  archivePrefix={arXiv},
  primaryClass={quant-ph}
}

@article{LeppajarviSedlak2021,
  title     = {Postprocessing of quantum instruments},
  author    = {Lepp\"{a}j\"{a}rvi, Leevi and Sedl\'{a}k, Michal},
  journal   = {Physical Review A},
  volume    = {103},
  number    = {2},
  pages     = {022615},
  year      = {2021},
  month     = {February},
  publisher = {American Physical Society},
  doi       = {10.1103/PhysRevA.103.022615},
  url       = {https://link.aps.org/doi/10.1103/PhysRevA.103.022615},
  archivePrefix = {arXiv},
  eprint    = {2010.15816},
  primaryClass  = {quant-ph}
}

@article{HorodeckiShorRuskai2003EntanglementBreaking,
  title         = {General Entanglement Breaking Channels},
  author        = {Horodecki, Michael and Shor, Peter W. and Ruskai, Mary Beth},
  journal       = {Reviews in Mathematical Physics},
  volume        = {15},
  pages         = {629--641},
  year          = {2003},
  doi           = {10.1142/S0129055X03001709},
  url           = {https://doi.org/10.1142/S0129055X03001709},
  eprint        = {quant-ph/0302031},
  archivePrefix = {arXiv},
  primaryClass  = {quant-ph},
  note          = {arXiv: https://arxiv.org/abs/quant-ph/0302031}
}

@article{Granade2012Robust,
  author        = {Granade, Christopher E. and Ferrie, Christopher
                   and Wiebe, Nathan and Cory, D. G.},
  title         = {Robust Online Hamiltonian Learning},
  journal       = {New Journal of Physics},
  volume        = {14},
  number        = {10},
  pages         = {103013},
  year          = {2012},
  doi           = {10.1088/1367-2630/14/10/103013},
  url           = {https://doi.org/10.1088/1367-2630/14/10/103013},
  eprint        = {1207.1655},
  archivePrefix = {arXiv},
  primaryClass  = {quant-ph}
}

@article{Wiebe2014Hamiltonian,
  author        = {Wiebe, Nathan and Granade, Christopher
                   and Ferrie, Christopher and Cory, D. G.},
  title         = {Hamiltonian Learning and Certification Using Quantum Resources},
  journal       = {Physical Review Letters},
  volume        = {112},
  pages         = {190501},
  year          = {2014},
  doi           = {10.1103/PhysRevLett.112.190501},
  url           = {https://doi.org/10.1103/PhysRevLett.112.190501},
  eprint        = {1309.0876},
  archivePrefix = {arXiv},
  primaryClass  = {quant-ph}
}

@article{Yu2023Robust,
  author        = {Yu, Wenjun and Sun, Jinzhao and Han, Zeyao and Yuan, Xiao},
  title         = {Robust and Efficient Hamiltonian Learning},
  journal       = {Quantum},
  volume        = {7},
  pages         = {1045},
  year          = {2023},
  doi           = {10.22331/q-2023-06-29-1045},
  url           = {https://doi.org/10.22331/q-2023-06-29-1045},
  eprint        = {2201.00190},
  archivePrefix = {arXiv},
  primaryClass  = {quant-ph}
}

@article{Bluhm2026Hamiltonian,
  author        = {Bluhm, Andreas and Caro, Matthias C. and Oufkir, Aadil},
  title         = {Hamiltonian Property Testing},
  journal       = {Quantum},
  volume        = {10},
  pages         = {1979},
  year          = {2026},
  doi           = {10.22331/q-2026-01-21-1979},
  url           = {https://doi.org/10.22331/q-2026-01-21-1979},
  eprint        = {2403.02968},
  archivePrefix = {arXiv},
  primaryClass  = {quant-ph}
}

@inproceedings{Gao2026Certification,
  author        = {Gao, Minbo and Ji, Zhengfeng and Wang, Qisheng
                   and Yu, Wenjun and Zhao, Qi},
  title         = {Quantum Hamiltonian Certification},
  booktitle     = {Proceedings of the 2026 Annual ACM-SIAM Symposium
                   on Discrete Algorithms (SODA)},
  pages         = {1424--1467},
  year          = {2026},
  publisher     = {Society for Industrial and Applied Mathematics},
  doi           = {10.1137/1.9781611978971.53},
  url           = {https://doi.org/10.1137/1.9781611978971.53},
  eprint        = {2505.13217},
  archivePrefix = {arXiv},
  primaryClass  = {quant-ph}
}

@article{Ji2026Barycentric,
  author        = {Ji, Kaiyuan and Mishra, Hemant K. and Mosonyi, Mil{\'a}n
                   and Wilde, Mark M.},
  title         = {Barycentric Bounds on the Error Exponents of Quantum
                   Hypothesis Exclusion},
  journal       = {IEEE Transactions on Information Theory},
  volume        = {72},
  number        = {4},
  pages         = {2391--2423},
  year          = {2026},
  month         = apr,
  doi           = {10.1109/TIT.2026.3661340},
  url           = {https://doi.org/10.1109/TIT.2026.3661340},
  eprint        = {2407.13728},
  archivePrefix = {arXiv},
  primaryClass  = {quant-ph},
  note          = {arXiv:2407.13728,
                   \url{https://arxiv.org/abs/2407.13728}}
}

@inproceedings{Ji2025Converse,
  author        = {Ji, Kaiyuan and Mishra, Hemant K. and Mosonyi, Mil{\'a}n
                   and Wilde, Mark M.},
  title         = {Converse Bounds for Quantum Hypothesis Exclusion:
                   A Divergence-Radius Approach},
  booktitle     = {2025 IEEE International Symposium on Information Theory
                   (ISIT)},
  pages         = {1--6},
  year          = {2025},
  publisher     = {IEEE},
  doi           = {10.1109/ISIT63088.2025.11195689},
  url           = {https://doi.org/10.1109/ISIT63088.2025.11195689},
  eprint        = {2501.09712},
  archivePrefix = {arXiv},
  primaryClass  = {quant-ph},
  note          = {arXiv:2501.09712,
                   \url{https://arxiv.org/abs/2501.09712}}
}

@article{Molmer2015Hypothesis,
  author        = {M{\o}lmer, Klaus},
  title         = {Hypothesis Testing with Open Quantum Systems},
  journal       = {Physical Review Letters},
  volume        = {114},
  pages         = {040401},
  year          = {2015},
  doi           = {10.1103/PhysRevLett.114.040401},
  eprint        = {1408.4568},
  archivePrefix = {arXiv},
  primaryClass  = {quant-ph}
}

@article{DattaFukudaHolevo2006,
  author  = {Datta, Nilanjana and Fukuda, Motohisa and Holevo, Alexander S.},
  title   = {Complementarity and Additivity for Covariant Channels},
  journal = {Quantum Information Processing},
  volume  = {5},
  pages   = {179--207},
  year    = {2006},
  doi     = {10.1007/s11128-006-0021-6}
}

@misc{FukudaHolevo2005,
  author        = {Fukuda, Motohisa and Holevo, Alexander S.},
  title         = {On Weyl-Covariant Channels},
  year          = {2005},
  eprint        = {quant-ph/0510148},
  archivePrefix = {arXiv},
  primaryClass  = {quant-ph}
}

@article{AudenaertScheel2008,
  author        = {Audenaert, Koenraad M. R. and Scheel, Stefan},
  title         = {On Random Unitary Channels},
  journal       = {New Journal of Physics},
  volume        = {10},
  pages         = {023011},
  year          = {2008},
  doi           = {10.1088/1367-2630/10/2/023011},
  eprint        = {0709.0824},
  archivePrefix = {arXiv},
  primaryClass  = {quant-ph}
}

@article{Nidhin2024SingleShotLabeling,
  title   = {Single-shot labeling of quantum observables},
  author  = {Sudarsanan Ragini, Nidhin and Ziman, M{\'a}rio},
  journal = {Physical Review A},
  volume  = {109},
  number  = {5},
  pages   = {052415},
  year    = {2024},
  month   = {may},
  doi     = {10.1103/PhysRevA.109.052415},
  eprint  = {2407.05351},
  archivePrefix = {arXiv},
  primaryClass  = {quant-ph}
}

@article{Arash2024MultipleShotLabeling,
  title         = {Multiple-shot labeling of quantum observables},
  author        = {Ghoreishi, Seyed Arash and Sudarsanan Ragini, Nidhin 
                   and Ziman, M{\'a}rio and Sazim, Sk},
  year          = {2024},
  eprint        = {2407.05392},
  archivePrefix = {arXiv},
  primaryClass  = {quant-ph},
  note          = {arXiv:2407.05392}
}

@article{Kretschmann2005QuantumMemory,
    author        = {Kretschmann, Dennis and Werner, Reinhard F.},
    title         = {Quantum Channels with Memory},
    journal       = {Physical Review A},
    volume        = {72},
    number        = {6},
    pages         = {062323},
    year          = {2005},
    doi           = {10.1103/PhysRevA.72.062323},
    eprint        = {quant-ph/0502106},
    archivePrefix = {arXiv},
    primaryClass  = {quant-ph}
}

@article{Heinosaari2017Incompatibility,
  author  = {Heinosaari, Teiko and Miyadera, Takayuki},
  title   = {Incompatibility of quantum channels},
  journal = {Journal of Physics A: Mathematical and Theoretical},
  volume  = {50},
  number  = {13},
  pages   = {135302},
  year    = {2017},
  doi     = {10.1088/1751-8121/aa5f6b},
  eprint  = {1608.01794},
  archivePrefix = {arXiv},
  primaryClass  = {quant-ph}
}

@misc{johnston2026sharpboundsperfectquantum,
      title={Sharp bounds for perfect quantum state classification beyond antidistinguishability}, 
      author={Nathaniel Johnston and Benjamin Lovitz and Vincent Russo and Jamie Sikora},
      year={2026},
      eprint={2609.17411},
      archivePrefix={arXiv},
      primaryClass={quant-ph},
      url={https://arxiv.org/abs/2609.17411}, 
}

@article{Ray2026EpistemicModels,
  title         = {Failure of epistemic models to explain
                   the antidistinguishability of quantum
                   mixed preparations},
  author        = {Ray, Sagnik and R, Visweshwaran
                   and Saha, Debashis},
  journal       = {Physical Review A},
  volume        = {113},
  number        = {6},
  pages         = {062203},
  year          = {2026},
  doi           = {10.1103/yhj8-qylw},
  eprint        = {2401.17980},
  archivePrefix = {arXiv},
  primaryClass  = {quant-ph},
  url           = {https://doi.org/10.1103/yhj8-qylw}
}

@misc{manna2026antidistinguishabilitystatesgeneralprobabilistic,
      title={Antidistinguishability of states in General Probabilistic Theories}, 
      author={Satyaki Manna and Anandamay Das Bhowmik},
      year={2026},
      eprint={2609.17498},
      archivePrefix={arXiv},
      primaryClass={quant-ph},
      url={https://arxiv.org/abs/2609.17498}, 
}

@misc{roy2026quantumstateexclusioncopies,
      title={Quantum state exclusion with many copies}, 
      author={Debanjan Roy and Tathagata Gupta and Pratik Ghosal and Samrat Sen and Somshubhro Bandyopadhyay},
      year={2026},
      eprint={2601.14410},
      archivePrefix={arXiv},
      primaryClass={quant-ph},
      url={https://arxiv.org/abs/2601.14410}, 
}

@book{busch2016quantum,
  title={Quantum measurement},
  author={Busch, Paul and Lahti, Pekka and Pellonp{\"a}{\"a}, Juha-Pekka and Ylinen, Kari and Lahti, Pekka Johannes},
  volume={23},
  year={2016},
  publisher={Springer, Berlin},
  nolink = {}
}
